\documentclass[12pt,preprint]{aastex63}
\usepackage{graphicx,graphics,aas_macros,amsmath}
\usepackage{color}
\usepackage{amssymb}
\usepackage{mathtools}
\usepackage{xspace}
\usepackage{rotating}
\usepackage{wrapfig}
\usepackage{appendix}
\usepackage{lipsum}
\usepackage{tikz,xcolor,hyperref}
\usepackage{lineno}

\newcommand{\chandra}{{\it Chandra}}

\newcommand{\suzaku}{{\it Suzaku}}
\newcommand{\xmm}{{\it XMM-Newton}}
\newcommand{\sax}{{\it BeppoSAX}}

\newcommand{\integral}{\textit{INTEGRAL}}
\newcommand{\nustar}{\textit{NuSTAR}}
\newcommand{\nicer}{\textit{NICER}}

\newcommand{\nh}{N$_{\rm H}$}
\newcommand{\nho}{NH$_{1}$}
\newcommand{\nht}{NH$_{2}$}

\newcommand{\ecyc}[1]{\ensuremath{E_{\rm{C}}}}
\newcommand{\src}{OAO 1657--415\xspace}

\definecolor{lime}{HTML}{A6CE39}
\DeclareRobustCommand{\orcidicon}{
	\begin{tikzpicture}
	\draw[lime, fill=lime] (0,0) 
	circle [radius=0.16] 
	node[white] {{\fontfamily{qag}\selectfont \tiny ID}};
	\draw[white, fill=white] (-0.0625,0.095) 
	circle [radius=0.007];
	\end{tikzpicture}
	\hspace{-2mm}
}
\foreach \x in {A, ..., Z}{\expandafter\xdef\csname orcid\x\endcsname{\noexpand\href{https://orcid.org/\csname orcidauthor\x\endcsname}
			{\noexpand\orcidicon}}
}

\newcommand{\ecut}{\ensuremath{E_{\rm{cut} }}}
\newcommand{\efold}{\ensuremath{E_{\rm{fold} }}}

\newcommand{\plcuteq}{\ensuremath{
  \mathrm{HIGHECUT}(E) = A\ E^{-\Gamma}\times\
  \begin{cases}
     1                           & (E\leq\ecut) \\
     {\rm e}^{-(E-\ecut)/\efold} & (E>\ecut)    \\
  \end{cases}
}}

 \newcommand{\newhcuteq}{\ensuremath{
  \mathrm{NEWHCUT}(E) = 
  \begin{cases}
  \ E^{-\Gamma}  &  E \leq \ecut - W/2 \\
  AE^3 + BE^2 + CE + D, &  \ecut - W/2 \leq E \leq \ecut + W/2 \\
   E^{-\Gamma}\times\ {\rm e}^{(\ecut - E)/\efold}, &  E \geq \ecut + W/2  \\
  \end{cases}
}}

\newcommand{\apjplcuteq}{\ensuremath{
   \mathrm{CUTOFFPL}(E) = A\ E^{-\Gamma}\times{\rm e}^{-(E)/\efold} \\
}}

\newcommand*{\ileyk}{\textcolor[rgb]{0,0.5,0}} 

\begin{document}

\title{Clumpy wind studies and the non-detection of cyclotron line in OAO 1657$-$415}

\author{\orcidA Pragati Pradhan}\thanks{pradhanp@erau.edu}
\affiliation{Massachusetts Institute of Technology, Kavli Institute for Astrophysics and Space Research, 70 Vassar St., Cambridge, MA, 02139, USA} 
\affiliation{Embry Riddle Aeronautical University, Department of Physics and Astronomy, 3700 Willow Creek Road, Prescott, AZ 86301, USA}

\author{\orcidB Carlo Ferrigno} 
\affiliation
{Department of astronomy, University of Geneva, chemin d'\'Ecogia, 16, CH-1290, Versoix, Switzerland}

\author{Biswajit Paul}
\affiliation{Raman Research Institute, Astronomy and Astrophysics, C. V. Raman Avenue, Bangalore 560080. Karnataka India}

\author{Enrico Bozzo}
\affiliation
{Department of astronomy, University of Geneva, chemin d'\'Ecogia, 16, CH-1290, Versoix, Switzerland}

\author{\orcidQ Ileyk El Mellah}
\affiliation{Institut de Planétologie et d'Astrophysique, UGA-CNRS, 414 Rue de la Piscine, 38400 Saint-Martin-d’Hères, France}

\author{\orcidF David P.\  Huenemoerder}
\affiliation{Massachusetts Institute of Technology, Kavli Institute for Astrophysics and Space Research, 70 Vassar St., Cambridge,
MA, 02139, USA}

\author{\orcidZ James F.\ Steiner}
\affiliation{Center for Astrophysics | Harvard \& Smithsonian, 60 Garden St., Cambridge, MA 02138, USA}

\author{\orcidV Victoria Grinberg}
\affiliation{European Space Agency (ESA), European Space Research and Technology Centre (ESTEC), Keplerlaan 1, 2201 AZ Noordwijk, The Netherlands}

\author{Felix Furst}
\affiliation{Quasar Science Resourced Ltd for ESA, European Space Astronomy Centre (ESAC), Camino Bajo del  Castillo s/n, 28692 Villanueva de la Ca\~nada, Madrid, Spain}

\author{\orcidR Chandreyee Maitra}
\affiliation{Max Planck Institute For Extraterrestrial Physics, 85748 Garching, Germany}

\author{\orcidP Patrizia Romano}
\affiliation{INAF, Osservatorio Astronomico di Brera, Via E.\ Bianchi 46, 23807 Merate, Italy}

\author{\orcidL Peter Kretschmar}
\affiliation{European Space Agency (ESA), European Space Astronomy Centre (ESAC), Camino Bajo del Castillo s/n, 28692 Villanueva de la Cañada, Madrid, Spain}

\author{\orcidH Jamie Kennea}
\affiliation{Department of Astronomy and Astrophysics, The Pennsylvania State University, 525 Davey Lab, University Park, PA 16802, USA}

\author{\orcidI Deepto Chakrabarty}
\affiliation{Massachusetts Institute of Technology, Kavli Institute for Astrophysics and Space Research, 70 Vassar St., Cambridge,
MA, 02139, USA}

\shorttitle{\nustar~and \nicer~results of OAO 1657$-$415}
\shortauthors{P.\ Pradhan et al.}

\begin{abstract}

Winds of massive stars are suspected to be inhomogeneous (or clumpy), which biases the measures of their mass loss rates. In High Mass X-ray Binaries (HMXBs), the compact object can be used as an orbiting X-ray point source to probe the wind and constrain its clumpiness. We perform spectro-timing analysis of the HMXB \src with non-simultaneous \nustar~and~\nicer~observations. We compute the hardness ratio from the energy-resolved light curves, and using an adaptive rebinning technique, we thus select appropriate time segments to search for rapid spectral variations on timescales of a few hundreds to thousands of seconds.  Column density and intensity of Iron K$\alpha$ line were strongly correlated, and the recorded spectral variations were consistent with accretion from a clumpy wind. We also illustrate a novel framework to measure clump sizes, masses in HMXBs more accurately based on absorption measurements and orbital parameters of the source. We then discuss the limitations posed by current X-ray spacecrafts in such measurements and present prospects with future X-ray missions. We find that the source pulse profiles show a moderate dependence on energy. We identify a previously undetected dip in the pulse profile visible throughout the \nustar~observation near spin phase 0.15 possibly caused by intrinsic changes in accretion geometry close to the neutron star. We do not find any evidence for the debated cyclotron line  at $\sim$ 36\,keV in the time-averaged or the phase-resolved spectra with \nustar.
\end{abstract}
\section{Introduction}
\label{intro}
 In High Mass X-ray Binaries (HMXBs), a compact object orbits around a massive hot star. Among them, Supergiant X-ray Binaries (SGXBs) host an O/B supergiant whose UV line-driven  wind is partly accreted onto the neutron star, which powers the X-ray emission. These winds are notoriously inhomogeneous, or `clumpy', although their clumpiness is still largely unconstrained (see \citealt{nunez17} for a recent review of accretion from inhomogeneous winds and \citealt{driessen2021,driessen2022} for detailed wind simulations). Changes in the mass accretion rate and absorption or scattering of X-rays lead to intrinsic and extrinsic variability respectively, with typical peak-to-trough X-ray intensity ratio of the order of 100 in classical SGXBs \citep{walter2015}. In some of these systems, accretion can also be quenched by magneto-centrifugal effects \citep{bozzo2008,ertan2017} and some of them are suspected to host a magnetar \citep{bozzo2021}. Provided we can disentangle between variability induced by accretion and that by magneto-centrifugal effects, the former offers a way to constrain wind clumpiness and the accretion mechanism.

In the extended neutron star magnetosphere, the magnetic field channels the accreted plasma onto the poles, forming accretion columns. The bulk of the X-ray emission observed from SGXBs is produced at the base of the accretion column, just above the neutron star surface (thermal mound). The photons of the soft (blackbody) emission originated in the mound are up scattered due to collisions with electrons in falling at high speed across the accretion column (bulk Comptonisation). Additionally, thermal Comptonisation occurs when high-energy photons lose energy, which produces an exponential cutoff. While the consensus is that the hard X-ray spectrum of SGXBs is produced by inverse Compton scattering of soft photons by hot electrons in the accretion column, the exact details (e.g., how X-rays interact with the highly magnetized plasma) are not fully understood. There are ongoing attempts to make new physical models for spectral formation process in X-ray pulsars based on this bulk and thermal Comptonisation of photons (see, e.g., \citealt{becker_wolf2007,kallman2015,west2020,caiazzo2021,sokolava2021}). The common practice to model the high energy emission from such systems therefore is to fit phenomenological multi-component models (e.g., power law with black body and high-energy cutoff) representing these physical processes. In addition, the X-ray spectra from these sources sometimes exhibit a Cyclotron Resonance Scattering Feature (CRSF). It is an absorption line produced by cyclotron resonant scattering of X-ray photons in the intense magnetic field near the surface, and it is the only direct diagnostics to measure the intrinsic magnetic field of the accreting neutron star.

\src is a SGXB situated at a distance of 7.1$\pm$1.3 kpc \citep{A06} that hosts 
an Ofpe/WN9 supergiant companion \citep[known for its large mass-loss rate;][]{MA09} and a 38 \rm{s} spinning neutron star, accreting from the stellar wind \citep{WP79}. With an orbital period of $\sim$10.4 d and an eccentricity of $\sim$ 0.1 \citep{C93}, the system occupies a distinct place in the Corbet diagram, which locates an accreting neutron star in HMXBs as a function of the neutron star spin period and the orbital period \citep{C86}. The source is located in between sources that  are thought to undergo mass transfer via Roche lobe overflow and those that accrete via stellar wind capture.  The pulsar in OAO 1657--415 exhibits dual accretion modes: one resembling direct wind accretion as in classical SGXBs, and the other accretion through a transient accretion disk \citep{B97,J12}. OAO 1657$-$415 is thus a fascinating system in a rare evolutionary state. Furthermore, the pulsar is completely eclipsed by the supergiant companion for $\sim$ 20 \% of the orbital period \citep{C93} and its light curve exhibits a remarkable flare-like X-ray variability that is typical of wind-fed systems \citep[]{BS08, pradhan2014}. 

The presence of a CRSF in \src has been a matter of debate. There has been a marginal detection of a cyclotron line at $\sim$ 36 keV in a \sax ~observation of the source, carried out during a moderate flux state \citep{O99}. This feature could not be confirmed in a dedicated \integral ~observation due to the limited energy resolution and statistics of the data, but it is worth noting that a depression around $\sim$ 49 keV was reported by \cite{BS08}. A dip-like feature at 32-34\,keV was also reported in \suzaku ~data, but its statistical significance was not high enough to firmly establish the presence of a cyclotron line \citep{pradhan2014,jaisawal2014}. \citet{jaisawal2021} studied data from ASTROSAT observations of \src, but did not investigate the presence of a possible cyclotron line, as their data {\bf were} limited to the 0.5--20\,keV band. 

We carried out this \nustar~observation with the aim of obtaining the source spectrum with better statistical quality than the previous observations in the hard X-ray band. Thanks to the large effective area, good spectral resolution and broadband coverage of \nustar, we can  investigate the presence of the debated CRSF feature in the source without any ambiguity. Two independent studies using the same  \nustar~(\emph{Nuclear Spectroscopic Telescope Array}) data, as in this work, were performed by different groups that claim cyclotron line detections at two slightly different energies \citep{saavedra2022,sharma2022}. Both groups reported model-dependent CRSFs with \citet{saavedra2022} reporting a CRSF at $\sim$ 28-38\,keV and a width of $\sim$ 10-25\,keV and \citet{sharma2022} reporting the CRSF at $\sim$ 38-42\,keV with width $\sim$ 0.8-5\,keV in the first two thirds of the observation. Since we have investigated the same data set, and are reporting on the same data, we provide the comparison of our findings with these studies later in this paper in section~\ref{sec:no-cyclo}.

Another interesting property of the X-ray emission from \src is related to the K$\alpha$ and K$\beta$ iron emission lines. Compared to other wind-fed systems, the strengths of these lines relative to the continuum are abnormally high even outside the eclipse (see, e.g., \citealt{pradhan2013,pradhan2015}). The equivalent width of these lines also proportionally scales with both the hardness ratio and the variable absorption column density \citep{pradhan2014}. The \chandra~spectrum of OAO 1657$-$415 clearly features ionized Hydrogen and Helium-like iron lines and a Compton shoulder, indicating the presence of a geometrically complex hot environment surrounding the X-ray source \citep{oao_hetg2019}. 

\begin{figure*}
\hspace{-1.cm}
\centering
\includegraphics[scale=0.6,angle=0]{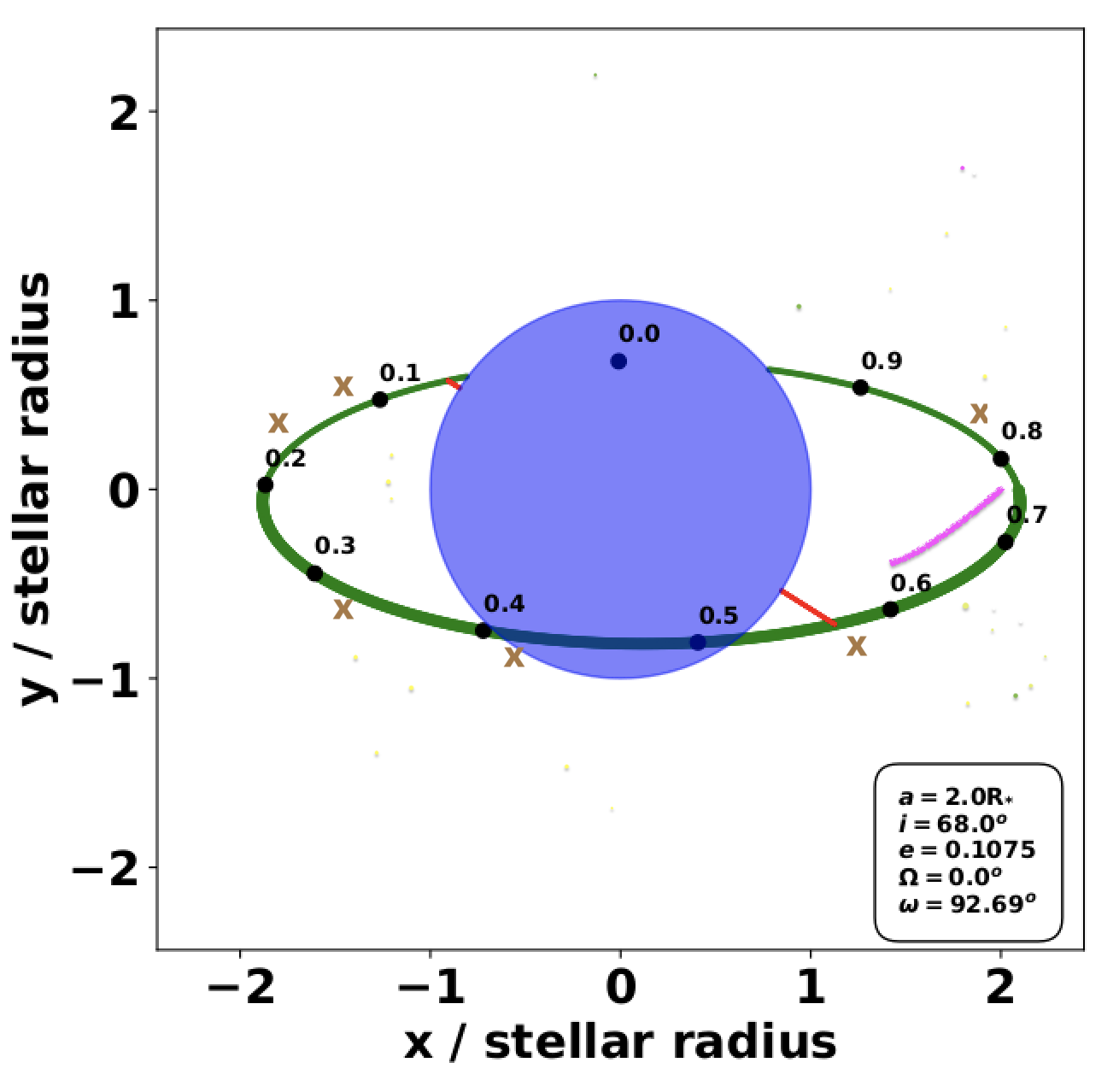}
    \caption{Orbit of the neutron star (solid green line) around its massive stellar companion (in blue) as seen from Earth. The values of orbital separation (a), the orbital inclination (i), eccentricity (e), longitude of periastron ($\Omega$), and periastron angle ($\omega$) are given in the lower right panel \citep{J12}. Black dots along the orbit mark orbital phases while the red solid line is the main axis of the orbit. The phases for \nustar~(magenta line) and \nicer~observations (brown `X' marks) are plotted.}
\label{orbit}
\end{figure*}

Neutron Star Interior Composition Explorer~(\nicer) is a payload onboard the International Space Station (ISS) launched in 2017 \citep{gendreau2016}. We also take advantage of the large effective area and observing agility of \nicer~to report high sensitivity monitoring of the source at different orbital phases. The details of the Observation IDs (ObsIDs) used in the paper along with the MJD, exposure and orbital phase of the observations are given in Table \ref{obsid}. In Fig.~\ref{orbit}, we mark the orbital phases relevant to these observations.

We discuss the details of the data reduction in Section \ref{data-red-nustar} and report the corresponding  results in Section \ref{results}. Our discussion is presented in Section \ref{disc} while a summary of the main findings in the paper is made available in Section \ref{summary}.

\section{Data reduction and analysis}
\label{data-red-nustar}
\subsection{\nustar}

\begin{table}
\caption{Table of observations used in the paper.}
\begin{tabular}{|c|c|c|c|c|}
\hline
Facility & ObsID & MJD & Duration (ks) & Orbital phase ($\phi$)  \\
\hline
\nustar & 30401019002 & 58645.536-58647.325 & 155 & 0.612-0.743  \\

\nicer & 3649010101 & 58961.359-58961.497  & 12 & 0.802-0.814  \\
 & 3649010301 & 58965.148-58965.350  & 18 & 0.165-0.184 \\
 & 3649010401 & 58967.665-58967.805 & 12 & 0.405-0.418 \\
 & 3649010501 & 58969.345-58969.483  & 12 & 0.566-0.579 \\
 & 3649010801 & 58975.093-58975.356  & 17 & 0.117-0.142 \\
 & 3649010901 & 58977.094-58977.356 & 23 &  0.308-0.333\\
 \hline
\end{tabular}
\label{obsid}
\end{table}

\src~was observed with \nustar~(\citealt{harrison2013}) in June 2019 during the orbital phase of 0.612-0.743, where phase 0 is the mid-eclipse time (MJD 52674.1199) and P$_{\rm orb}$ = 10.447355\,d \citep{falanga2015-orbit} that we use to determine orbital phase throughout the paper. \nustar~consists of two focal plane modules, FPMA and FPMB, each made of four pixelated detectors (DET0-DET3) spanning an energy range of 3--79\, keV. We used \texttt{nupipeline} version 0.4.6 to generate cleaned event files, which also provides the recommended source and background regions. These region files were an input to \texttt{nuproducts} that we used for extracting data products like light curves and spectrum.

We fit the FPMA and FPMB spectra simultaneously (keeping the relative instrument normalization free)  using \texttt{HEASOFT} 6.26.1 and \texttt{CALDB} version 20201217 with \texttt{XSPEC} 12.10.1f. The spectra were binned using an optimal binning scheme adapted from \citet{kaastra2016}. We performed timing analysis in the barycenter-corrected frame and corrected the \nustar~light curves for orbital motion and subtracted the background. 
Using the epoch folding technique, we determined the pulse period, which was used to create the energy resolved pulse profiles of OAO 1657$-$415 and to obtain the pulse phase resolved spectra (details in Section \ref{results}).




\subsection{NICER}
\label{data-red-nicer}
The \nicer~observations of \src, although not simultaneous with the \nustar~observations,  are very useful for studying the variation of absorption with orbital phase. We monitored \src~with \nicer~during April-May 2020 (ObsID: 364901090) for three orbital cycles of the binary (see Table \ref{obsid}). For \nicer~data reduction, screening has been applied to all NICER data using the calibration release {\sc xti20200722} as follows: Data from detectors 14, 34, and 54 were excised, as these detectors have exhibited intermittent problems being prone to excess electronic noise (e.g., \citealt{riley2019}). Good time intervals are defined using continuous intervals with no packet loss, in which the ISS was outside the South Atlantic Anomaly and pointing aligned within 1.5 arcmin to the target, with all subsystems configured nominally.  We require SUN\_ANGLE $>$ 60 degrees (or when in shadow, SUNSHINE==0), with ELV $>$ 20 and BR\_EARTH $>$ 30.  In addition, the remaining 49 active detectors were screened based on X-ray overshoot, and undershoot rates which serve as indicators of liveness, particle background, and optical loading, respectively.  Any detector outside the median by > 15 median absolute deviation (MAD) (roughly, an unbiased equivalent of $\approx$ 10$\sigma$) is rejected. Calibration ARF and RMF files were generated, respectively, by summing or averaging the per$-$FPM products. 
In selecting good events, the default fast-to-slow ratio (`trumpet') filter was employed, and the background spectra were derived using the standard `3C50' background model \citep{remillard2021}. 


For the same filtering criteria as described above, the \nicer~light curves were extracted from the event files obtained from \texttt{nicerl2} to which barycentric corrections were performed with \texttt{barycorr} using latest ephemeris file JPLEPH.430. We then searched for pulsations in each ObsIDs using \texttt{efsearch} and generated pulse profiles using \texttt{efold}.

\section{Results}
\label{results}
\subsection{\nustar~results}
The \nustar~analysis thread recommend fitting the FPMA and FPMB spectra separately while keeping their relative instrument normalization free. The spectra for \src~are shown in Fig.~\ref{nustar_spectrum}. Since the spectra above 70\,keV are background dominated, we fit the 3--70\,keV \nustar~spectra with  phenomenological \texttt{XSPEC} \citep{xspec1996} continuum models {\it viz.,} CUTOFFPL, HIGHECUT \citep{white1983} and NEWHCUT \citep{burderi2000}. 

\begin{equation}
\apjplcuteq   
\end{equation}

\begin{equation}
  \plcuteq  
\end{equation}

\begin{equation}
\newhcuteq 
\end{equation}

While the fits were comparable using these models (i.e., similar ${\chi_{r}^{2}}$ per degrees of freedom; d.o.f), the MCMC corner plots\footnote{especially for time-resolved and phase-resolved spectra.} indicated that CUTOFFPL permits the least degeneracy among the spectral parameters. Therefore, the best continuum to describe the source spectrum corresponds to an absorbed power law with an exponential cutoff and a Gaussian function to model the K$_\alpha$ emission line leading to ${\chi_{r}^{2}}$/d.o.f.\ as 1.72/666. With this model, we still see some wavy residuals (especially below 5\,keV) and in order to smoothen it, we added a partial covering model to account for local absorption which improved the ${\chi_{r}^{2}}$/d.o.f.\ to 1.24/665 d.o.f.\ This inclusion of two absorption components is motivated by the fact that absorption column density in HMXBs (\nh) comprise two components: global absorption, \nho,~that arise from a component that is extended, possibly the stellar wind, large-scale
structures like an accretion wake \ileyk{\citep{mellah2015}}, Co-rotating Interaction Regions {\citep[CIRs,][]{naze2013,massa2014}}, and the interstellar column density \citep{wilms2000}. The second component is for local absorption \nht~very close to the neutron star that blocks the compact source partially (with partial covering fraction \rm{f}) in size or partially in time (clumps/structures). 

With the addition of a partial covering model,  the value of absorption column density,  \nho~could not be constrained. We therefore froze this to the Galactic line of sight absorption of 1.8 $\times$ 10$^{22}$ atoms cm$^{-2}$ and the free component, \nht,~encompasses a time average of the local absorption, which is not fully covering due to moving structures/clumps.
The latter component of the partial covering model is well known to be a crucial component necessary to fit the X-ray spectrum of \src \citep{A06,pradhan2014,jaisawal2014,jaisawal2021,oao_hetg2019}. 

\begin{figure*}
\centering
\includegraphics[height=15cm,width=12cm,angle=-90]{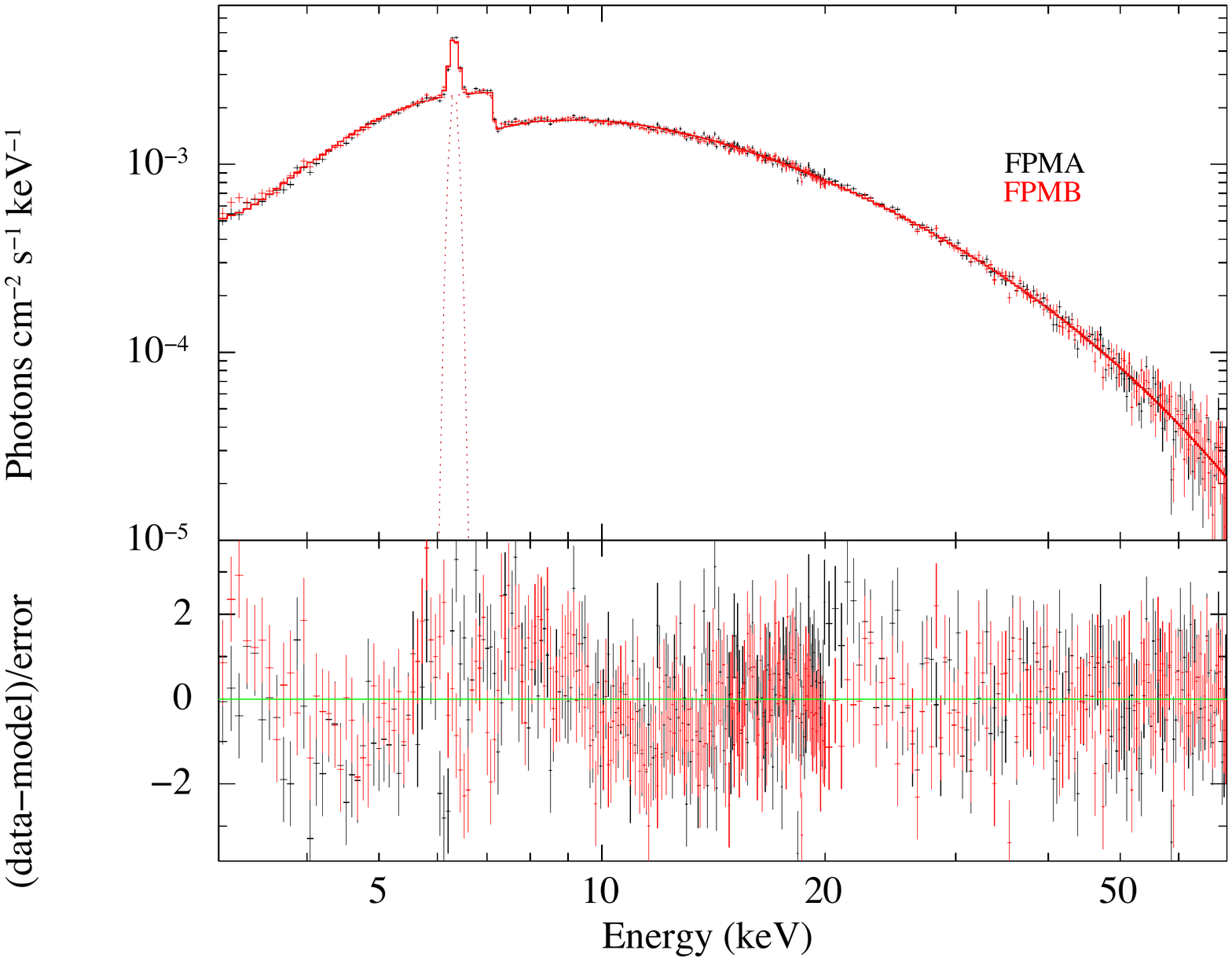}

\vspace{2.5cm}
\caption{Time-averaged spectrum of \src for \nustar~FPMA (black), FPMB (red) modelled with CUTOFFPL for orbital phase {0.612--0.743}. The corner plot for spectral parameters for the time averaged spectrum is given in Fig.~\ref{corner:avg}. }
\label{time-average}
\end{figure*}

Note that, in addition to the K$_\alpha$ line, since the X-ray spectrum of \src~ exhibits complex iron line features {\it viz.} Fe XXVI (6.97\,keV) and Fe XXV (6.7\,keV)  and K$_\beta$ (7.1\,keV), we also tried to account for these lines by freezing the line energy and width of these lines detected with Chandra/HETG \citep{oao_hetg2019} while allowing the normalization to vary. However, even with this exercise, we were not able to constrain the normalization of these lines – possibly due to the limited spectral resolution of \nustar~- nor was there any improvement in the spectral fitting. The \nustar~spectrum therefore only allowed constraining the Gaussian parameters for the Fe K$_\alpha$ line in the source. 
We list the complete parameters for spectral fitting in Table \ref{nustar_spectrum} and the time-averaged spectrum is shown in Fig.~\ref{time-average}. The MCMC corner plot for the average spectrum is shown in Fig.~\ref{corner:avg}. 

X-ray variability of HMXBs can be studied using a few different techniques. For instance, the tracks exhibited by wind-accretion HMXBs on color-color diagram during absorption can be explained by partial covering model with variable absorption \citep{hirsch2019,grinberg2020}. Another technique is to investigate the X-ray variability through hardness-ratio (hard/soft) resolved spectral analysis: since soft X-ray photons are more easily absorbed than higher energy photons, the X-ray variability caused by absorption is reflected in the hardness-ratio (hard/soft) resolved spectral analysis (see, e.g., \citealt{grinberg2017A_vela,pradhan2019_igrj18027}).

\begin{figure*}
\centering
\includegraphics[scale=0.33,angle=0]{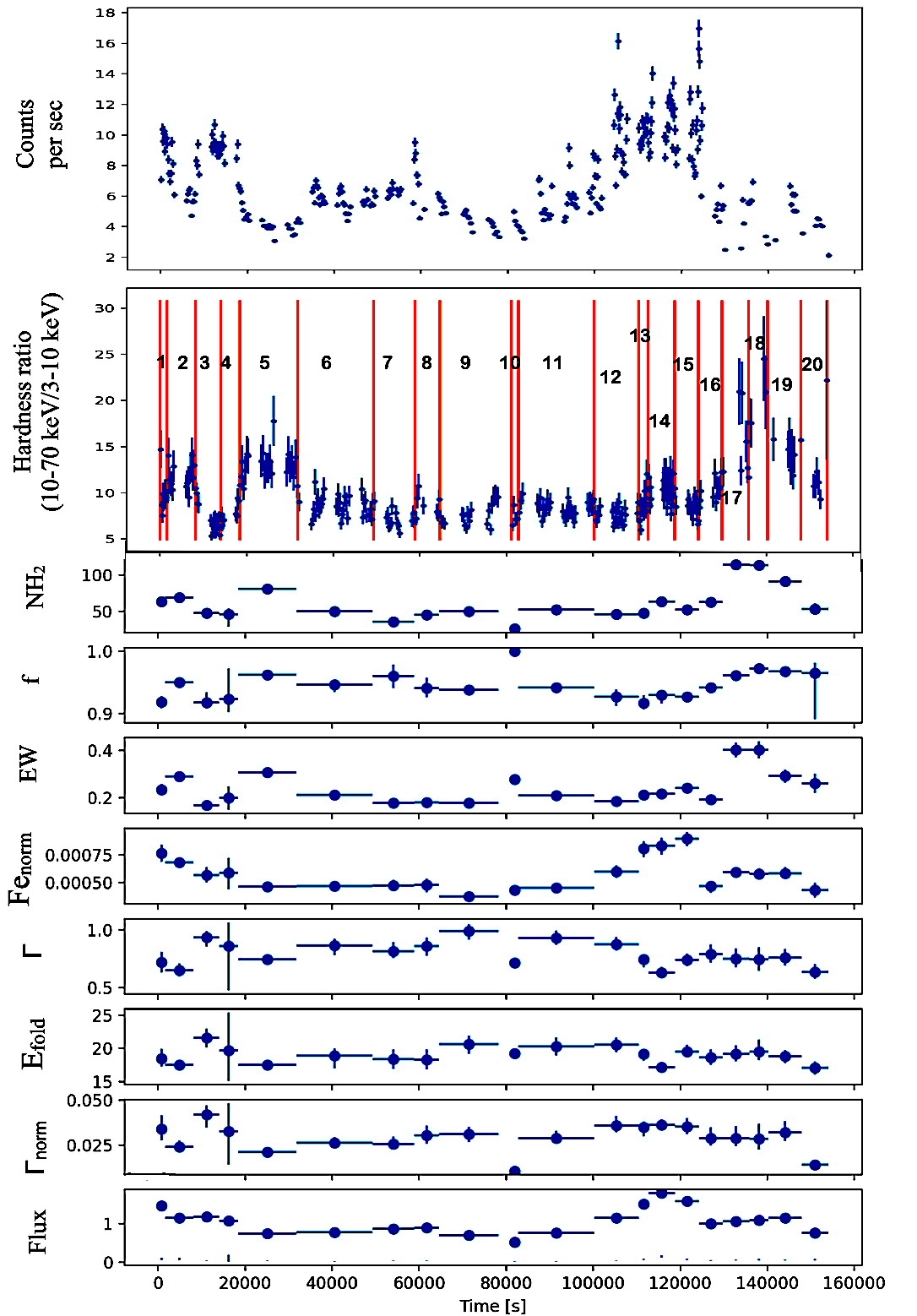}
\caption{From top to bottom are the \nustar~light curves (counts per second), Hardness-ratio (HR) with 20 time segments delimited by red vertical solid lines, absorption column density N$_{\rm H2}$ (in units of 10$^{22}$ cm$^{-2}$), covering fraction f, equivalent width (EW in keV), normalization or Fe K$\alpha$ line Fe$_{\rm{norm}}$ (in photons cm$^{-2}$ s$^{-1}$), spectral index $\Gamma$, folding energy E$_{\rm{fold}}$ (in keV), normalization of the power law $\Gamma_{\rm{norm}}$  (in photons keV$^{-1}$ cm$^{-2}$ s$^{-1}$ at 1 keV) and absorbed flux in 3--70\,keV  (in 1$\times~10^{-9}$ erg cm$^{-2}$ s$^{-1}$). N$_{\rm H1}$ is frozen at the line of sight absorption of $1.8\times 10^{22}$ atoms cm$^{-2}$. Note that not all parameters could be constrained in segment 10 and therefore no errors are plotted for this segment. The corner plots corresponding to some segments where the absorption values are medium, low, and high (1, 7, 19) in these fits are shown in Figures~\ref{corner:hr1}, \ref{corner:hr2} and \ref{corner:hr3}.}
\label{variation-time-res}
\end{figure*}

In this paper, we have {performed hardness-ratio (HR; 10-70\,keV/3-10\,keV) resolved spectral and timing analysis}: we determined the time segments where the hardness ratio changed significantly by using the Bayesian Blocks analysis \citep{bayesianblocks}. Such a technique has also been previously used in the case of supergiant fast X-ray transients, SFXTs by \citet{sidoli19} to identify intervals of significant intensity variability in the light curves of a sample of these objects observed by \xmm. The block fitness function used is the one for point measurements and the \texttt{ncp\_prior} was chosen to have a false alarm rate around 1\% using Fig.~6 of \citet{bayesianblocks}. We verified that the number of spurious changes of the HR was compatible with expectations by running a sample of 50 simulations with constant count rate, drawn from a distribution with noise equivalent to the measured light curve of each source. We obtained 21 time intervals with this technique where the HR is revealed to undergo significant variations. When we fit the HR resolved spectra from these 21 segments for the two FPMs, we find that one of the segments showed a possible degeneracy in the spectral parameters and therefore to mitigate any statistical limitation, we merged this segment with the next immediate one. As a result, we therefore divided the whole observation in 20 segments and the fits to 
these HR-resolved segments (shown in Fig.~\ref{variation-time-res}) were performed using C-statistics and the confidence intervals were determined using MCMC chains.

In Fig.~\ref{variation-time-res}, we show the \nustar~light curves (counts per second), hardness-ratio (HR) with 20 time segments delimited by red vertical solid lines, local absorption column density \nht~(in units of 10$^{22}$ cm$^{-2}$), covering fraction f, equivalent width (EW in keV), normalization or Fe K$\alpha$ line (Fe$_{\rm{norm}}$ in photons cm$^{-2}$ s$^{-1}$), spectral index $\Gamma$, folding energy E$_{\rm{fold}}$ (in keV), normalization of spectral index ($\Gamma_{\rm{norm}}$ in photons keV$^{-1}$ cm$^{-2}$ s$^{-1}$ at 1 keV) and absorbed flux in 3--70\,keV  (in units of 1$\times~10^{-9}$ erg cm$^{-2}$ s$^{-1}$). From the plot we make the following observations (i) The HR and \nht~to follow a similar variability trend, (ii) The \nht~and EW of Fe line are correlated, (iii) The covering fraction always remains high, ranging between 90-100\%. The corner plots corresponding to some segments in Fig.~\ref{variation-time-res}, where the absorption values are medium, low, and high (1, 7, 19) in these fits are shown in Figures~\ref{corner:hr1}, \ref{corner:hr2} and \ref{corner:hr3}.

\begin{figure}
\hspace{-2.0cm}
\includegraphics[height=8.5cm, width=7.5cm, angle=-90]{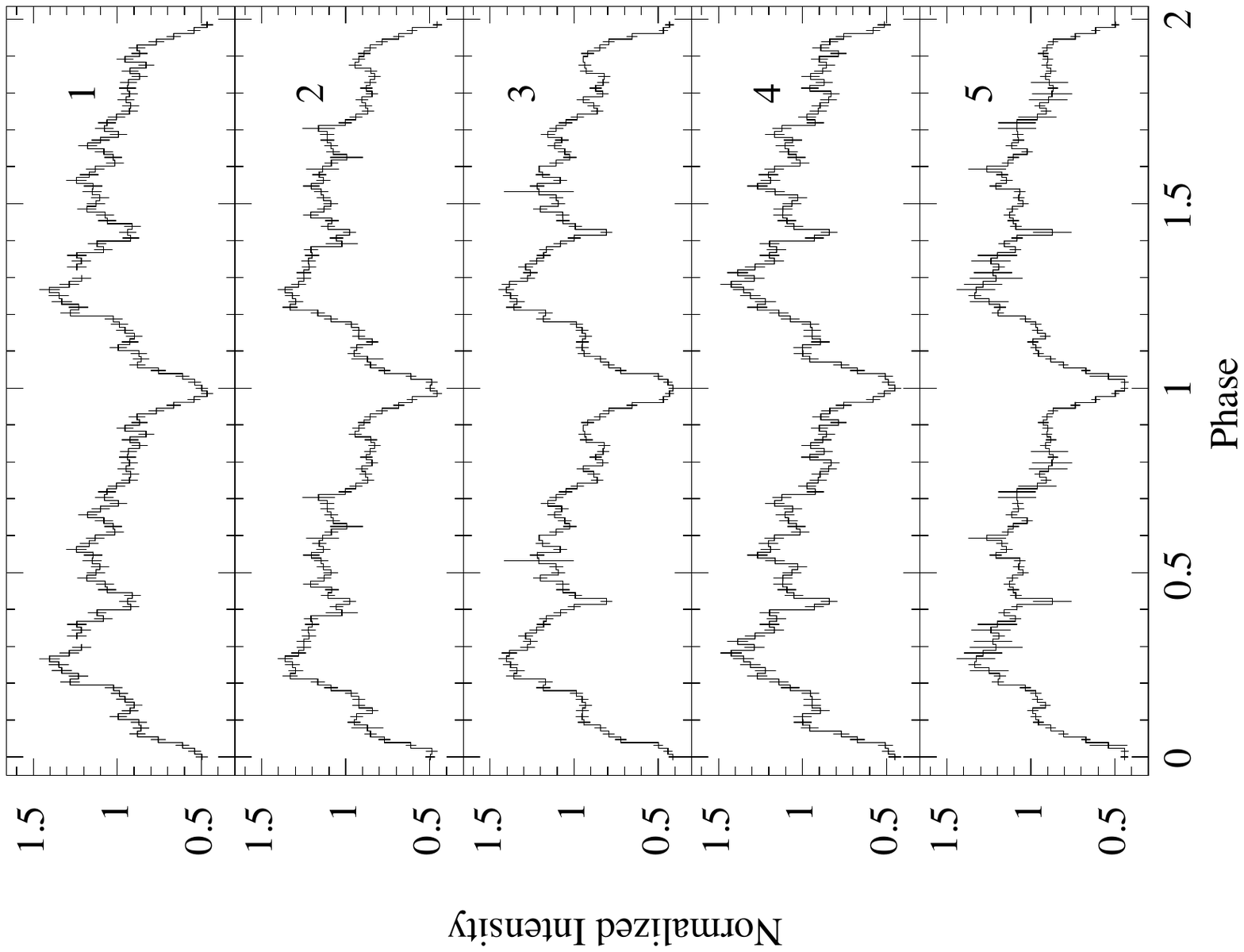}
\hspace{-5cm}
\includegraphics[height=8.5cm, width=7.5cm,angle=-90]{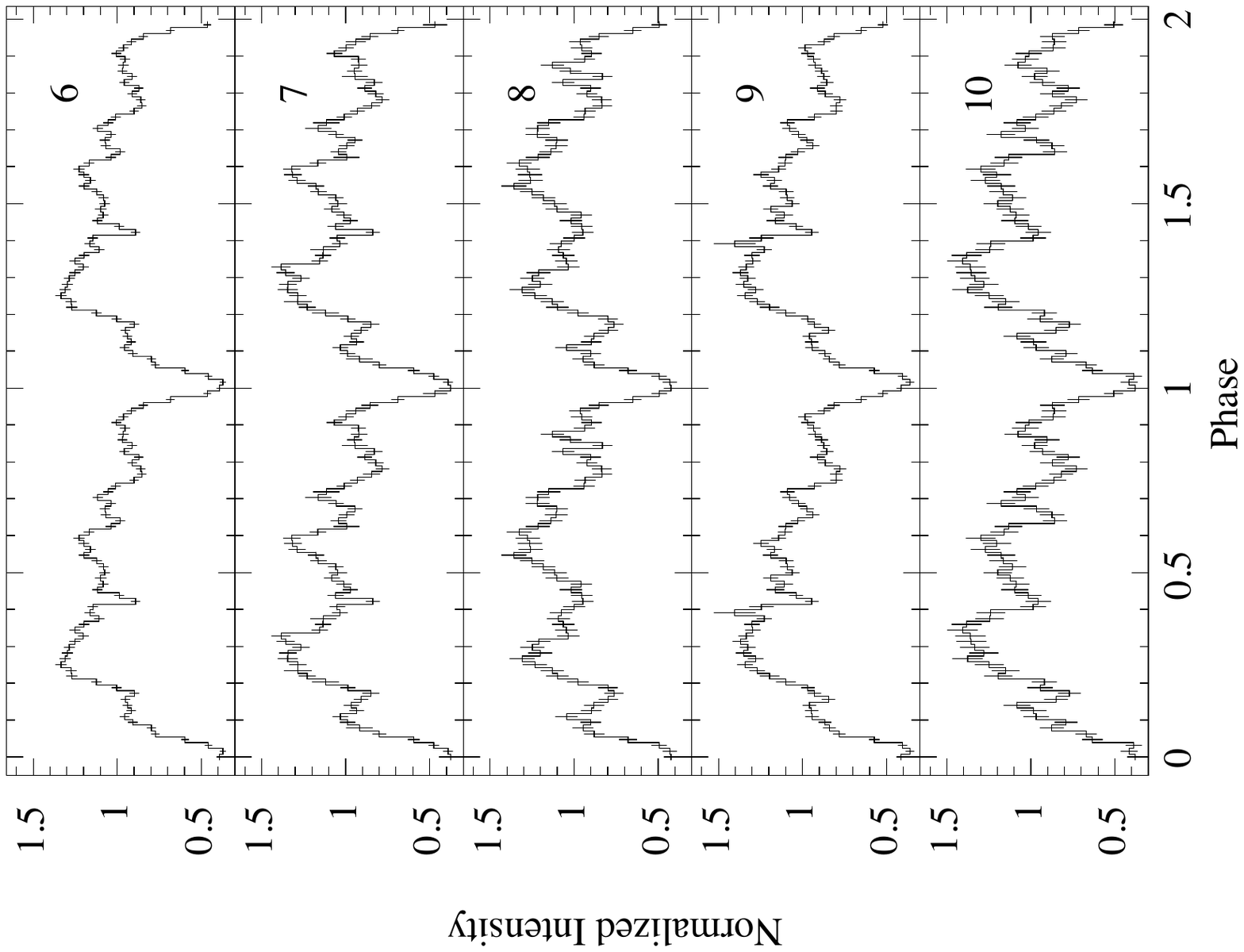}
\hspace{-5cm}
\includegraphics[height=8.5cm, width=7.5cm,angle=-90]{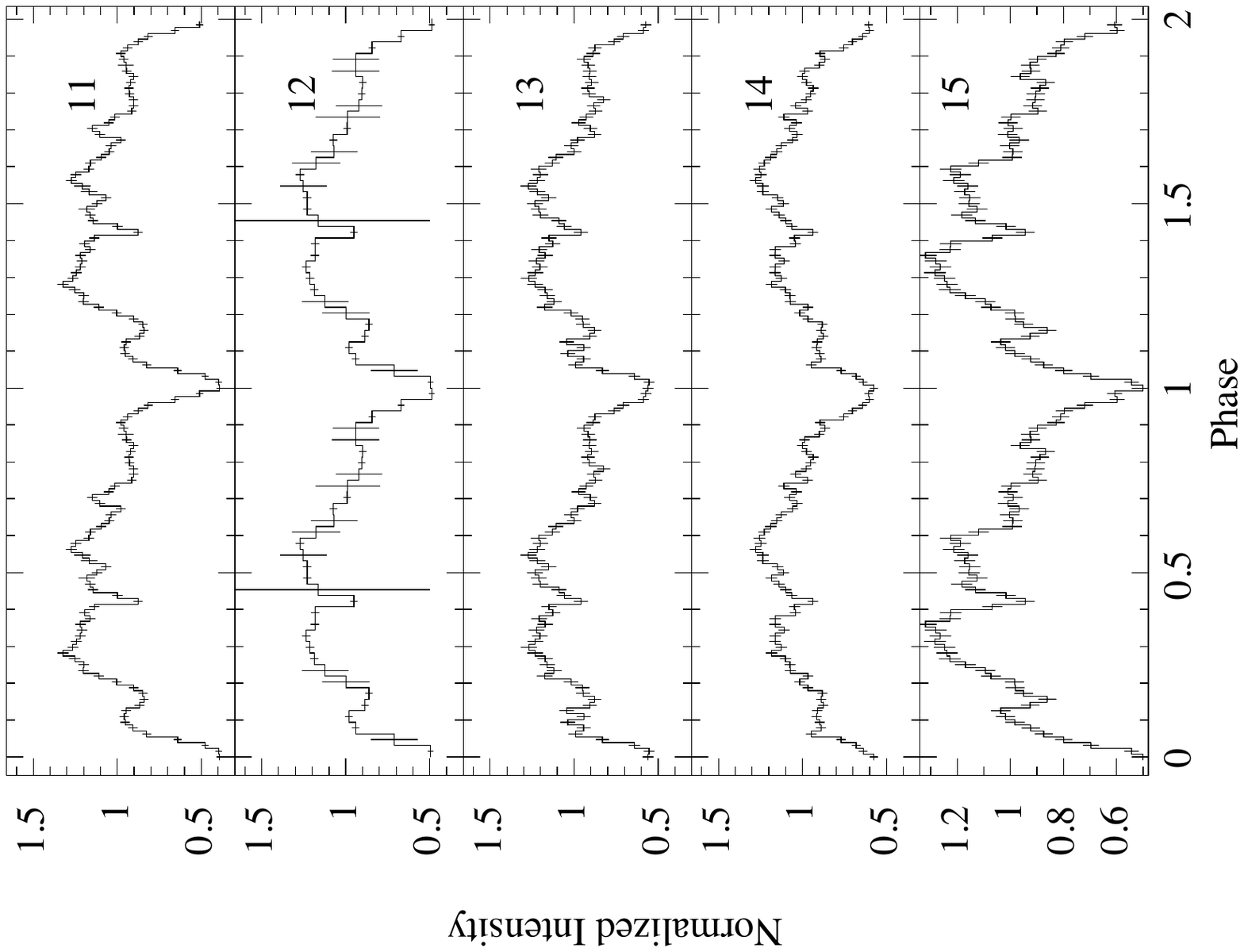}
\hspace{-5cm}
\includegraphics[height=8.5cm, width=7.5cm,angle=-90]{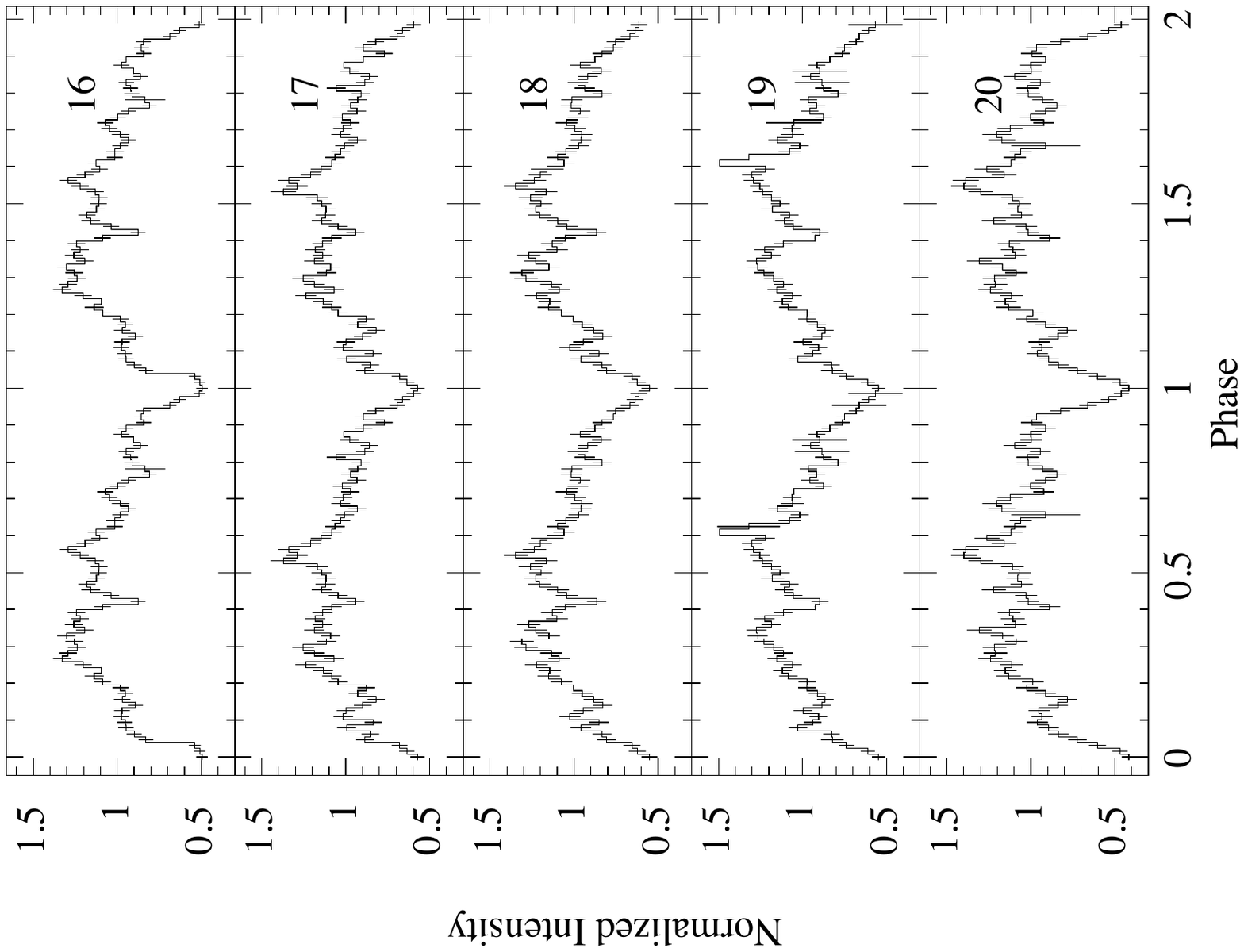}

\vspace{1.5cm}
\caption{Pulse profiles over each time segment of constant hardness ratio (see panel in Fig.~\ref{variation-time-res}).
Time segments are indicated in the upper right corner. The Y-axis is the normalized counts/sec with respect to the average for each plot.}
\label{pp}
\end{figure}

We extracted the pulse profiles for each of these 20 time segments (Fig.~\ref{pp}) and investigated the pulse profile morphology with time. For pulse profile analysis, we performed the orbital correction (using orbital parameters from \citealt{falanga2015-orbit}) and we use \texttt{efsearch} to determine the pulse period of the pulsar to be 37.01072\,s ($\pm$ 0.00034) at MJD 58645. 

The pulse profiles are created from the orbital and barycenter corrected light curve with the above period and an epoch of MJD 58645.0000857 so that the minimum of the first peak i.e., the peak more prominent at high energies, is approximately at pulse phase zero. Our choice coincides with the phase zero in \citet{sharma2022} and can be used to compare the results obtained from many other satellites since it is a very recognizable feature in the pulse profile of \src. Note that for clarity, we show pulse profiles twice over the same spin phase.

\begin{figure}
\includegraphics[scale=0.7,angle=0]{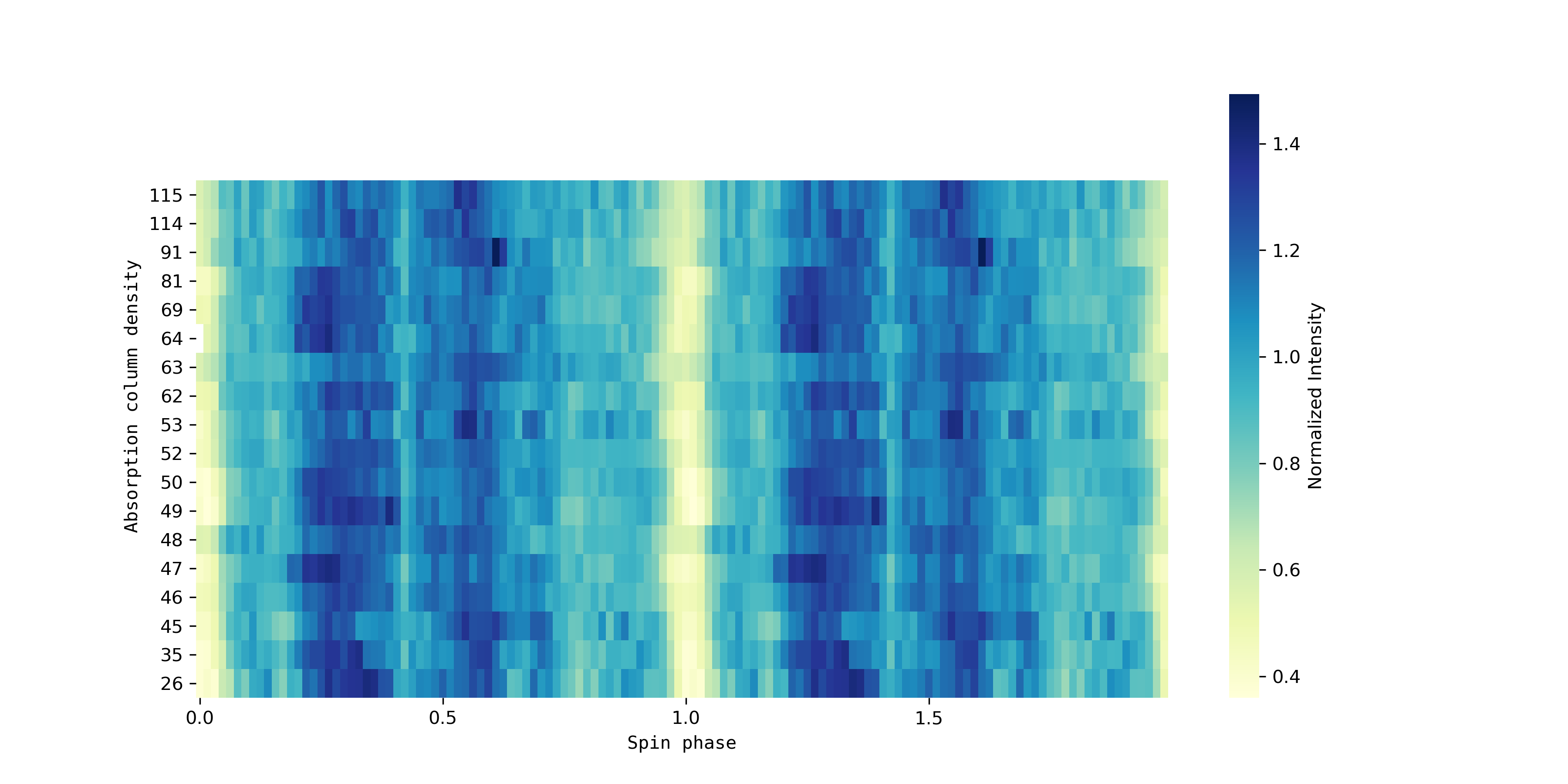}
\caption{Pulse profiles of \src for HR-resolved segments in order of increasing absorption column density (units of 10$^{22}$ cm$^{-2}$) marked along the Y-axis from down upwards and spin phase along the X-axis. The color maps represent the pulse normalized intensity.}
\label{pp-heatmap}
\end{figure}

The same pulse profiles evolution with time and absorption are shown in Fig.~\ref{pp} and Fig.~\ref{pp-heatmap} respectively. From these two figures, we can see no significant evolution of the pulse profiles with either time or absorption.

\begin{figure*}
\centering
\hspace{-3cm}
\includegraphics[scale=0.47,angle=-90]{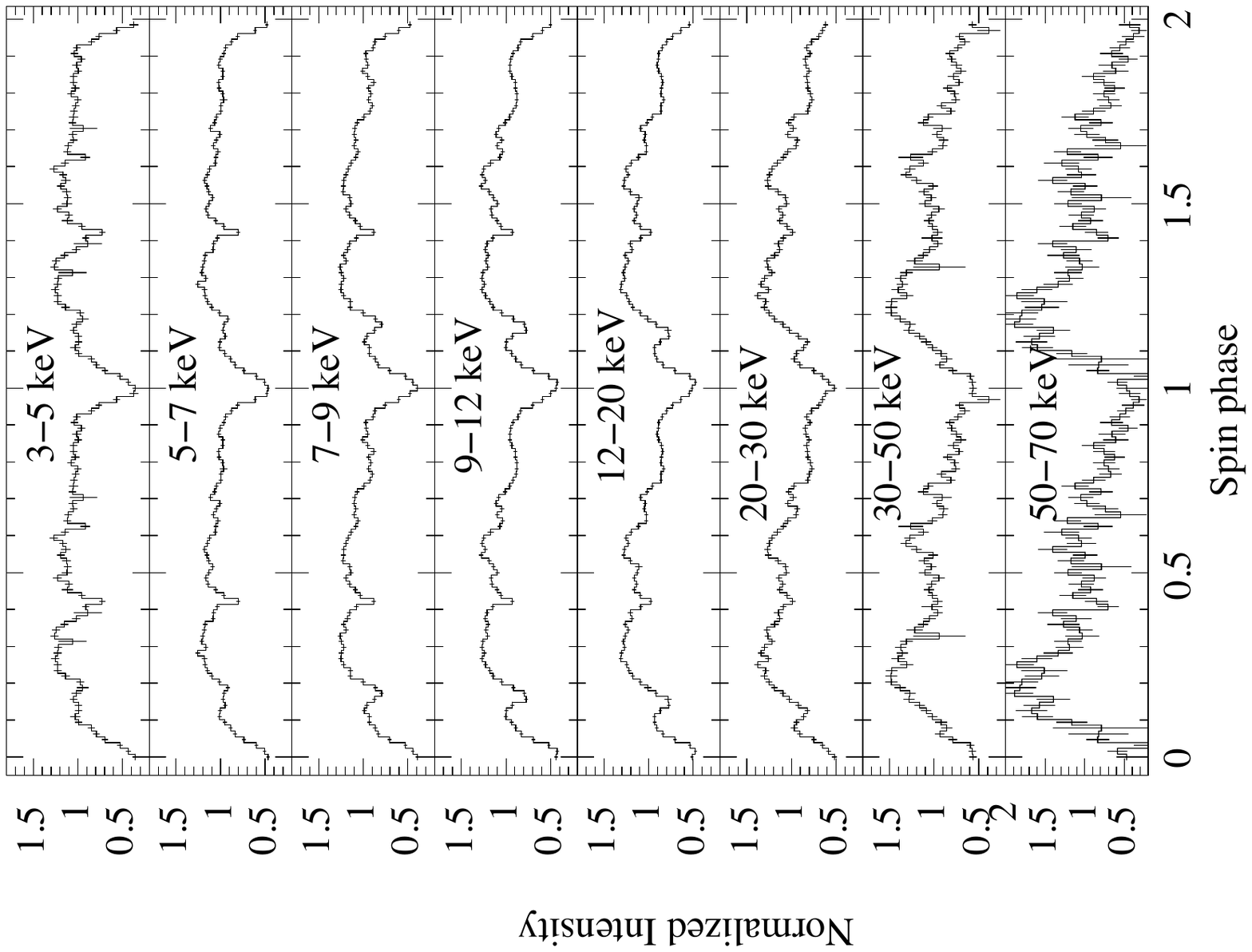}
\hspace{-7.5cm}
\includegraphics[scale=0.47,angle=-90]{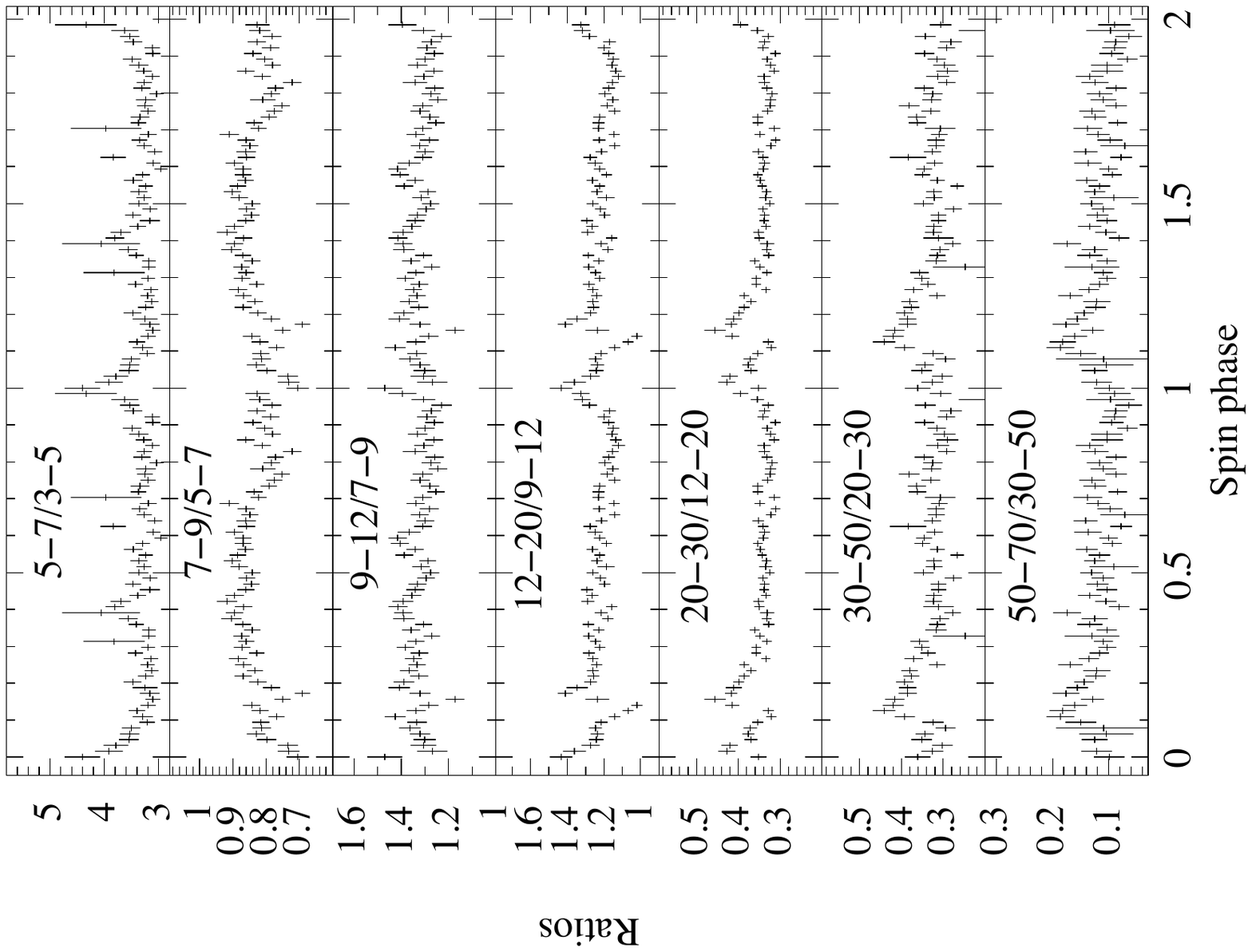}

\vspace{2.5cm}
\caption{Left: Energy-resolved pulse profiles for FPMA+B light curves where a dip at $\sim$ 0.15 is clearly seen. Right: Ratio of pulse profiles with energy denoted inside the panel.} 
\label{energy-pp}
\end{figure*}

Furthermore, in order to investigate the variability of pulse profiles with energy, we also extracted the pulse profiles for various energy ranges shown. In some HMXBs, the energy dependence of pulse profiles is not prominent (e.g., \citealt{pradhan2015}). Thanks to the large area of \nustar, we were able to identify a dip near spin phase 0.15 for the first time in \src. We show the energy-resolved pulse profiles on the left of Fig.~\ref{energy-pp} and the ratio to the right, to find that (i) the dip at spin phase 0.15 seems to be evolving with energy and increase from 5-30\,keV and vanishes thereafter, and (ii) dips at spin phases 0, 0.4 and 0.6 become less prominent with increasing X-ray energy. 

\begin{figure*}
\flushleft
\includegraphics[scale=0.65, angle=-90]{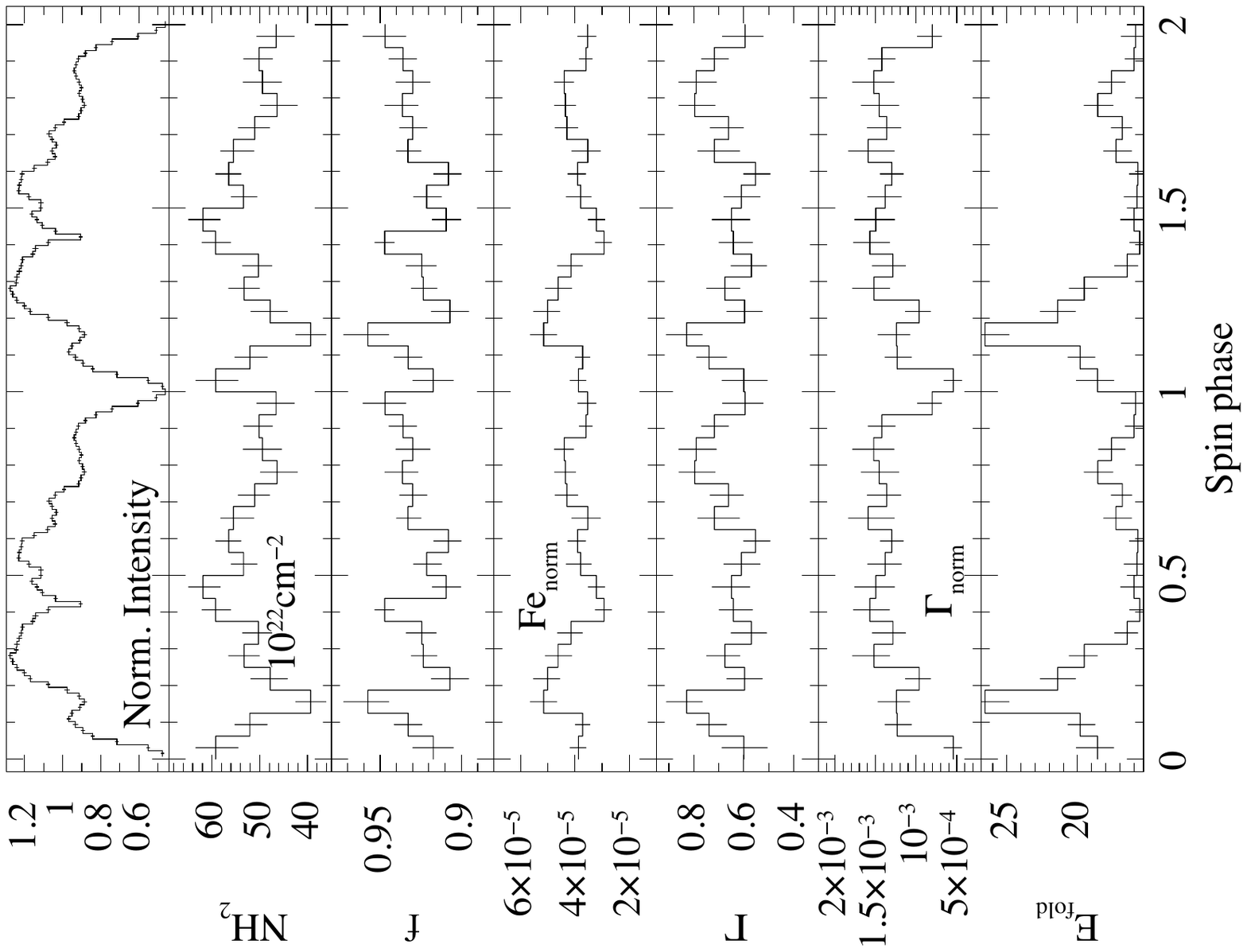}
\caption{Spin-phase resolved NuSTAR spectroscopy of \src for 16 phase-bins. From top to bottom are the normalized intensity of pulse profiles, absorption column density \nht~(in units of 10$^{22}$ cm$^{-2}$), covering fraction f, Fe K$\alpha$ normalization (Fe$_{\rm{norm}}$; in photons cm$^{-2}$ s$^{-1}$), spectral index $\Gamma$, followed by normalization of the power law ($\Gamma_{\rm{norm}}$; in photons keV$^{-1}$ cm$^{-2}$ s$^{-1}$ at 1 keV) and folding energy (E$_{\rm{fold}}$; in keV). The corner plots of the third, seventh, and thirteenth phase-bins, where the absorbing column density is low, high, and medium, respectively, are shown in  Figures~\ref{corner:pr1}, \ref{corner:pr2}, and \ref{corner:pr3}.
}

\label{spec-phase}
\end{figure*}

Finally, we performed spin-phase resolved spectral analysis in 16 phase bins using the same model as determined by the average spectrum. The average pulse profile is plotted on the top of Fig.~\ref{spec-phase} and the absorption on the second panel from the top. The absorption increases during dips in the pulse profiles, except during the dip around 0.15 where there is a decrease in local absorption  \nht~and changes in the folding energy, E$_{\rm{fold}}$. We see a variation of the Fe K$\alpha$ normalization (Fe${_{\rm norm}}$; in photons cm$^{-2}$ s$^{-1}$) with spin phase. The covering fraction $\rm{f}$ does not change much with phase, and the normalization of the power law ($\Gamma_{\rm{norm}}$; in photons keV$^{-1}$ cm$^{-2}$ s$^{-1}$ at 1 keV) scales with X-ray normalized intensity for each segment. The Pearson coefficient between the count rate (proxy for normalized intensity) and $\Gamma_{\rm{norm}}$ for the 16 segments is $\sim$ 0.92. All these spectral variations remain valid following the inspection of corner plots as well. For reference, we show corner plots in Figures~\ref{corner:pr1}, \ref{corner:pr2}, and \ref{corner:pr3} of three-phase intervals where the absorbing column density is low, high, and medium.

\subsection{\nicer~results}

\begin{figure}
\begin{tabular}{cccc}
\hspace{-2.0cm}
\includegraphics[height=1.5in,width = 1.8in,angle=0]{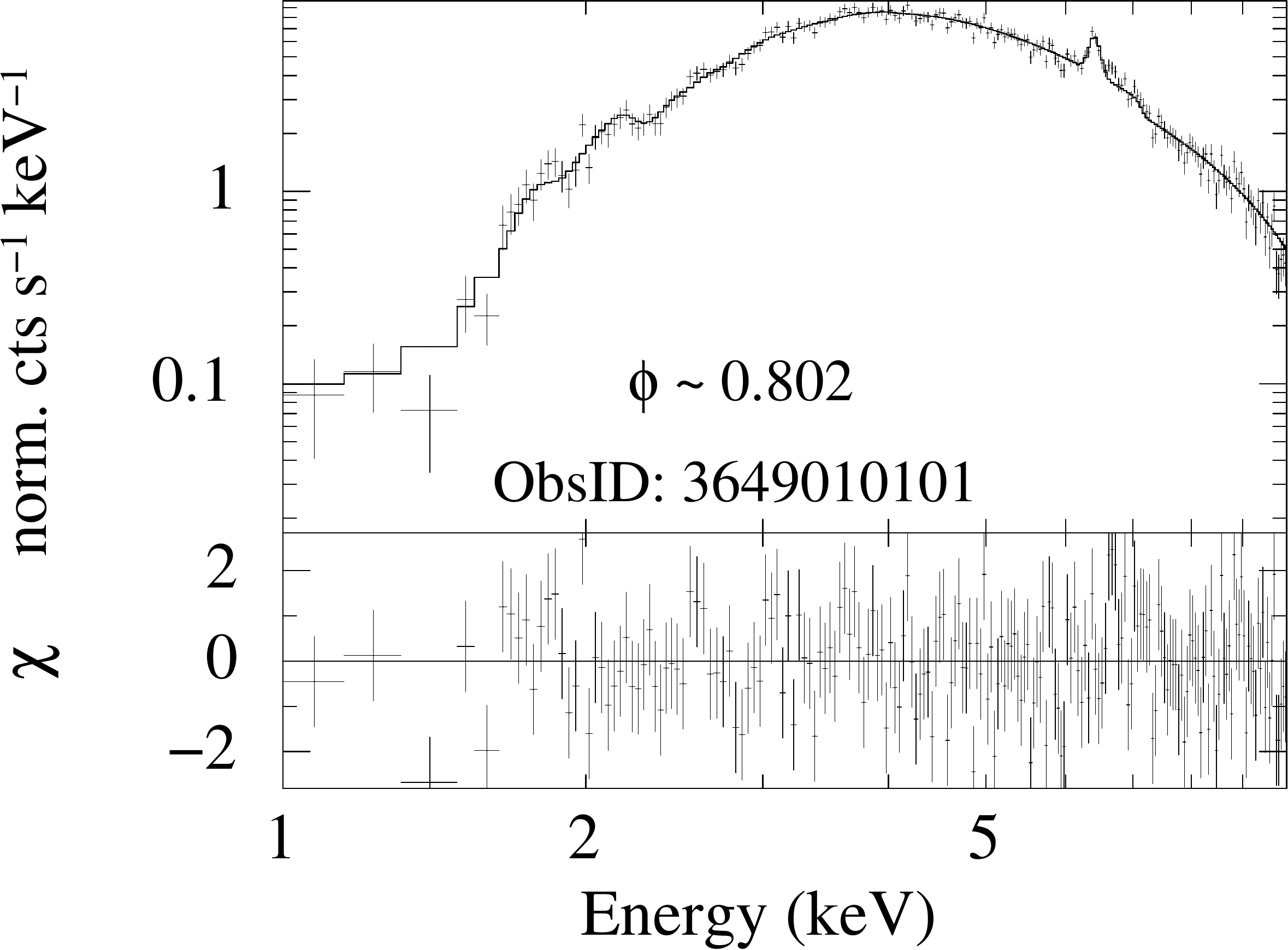} &
\hspace{-0.7cm}
\includegraphics[height=1.5in,width = 1.8in,angle=0]{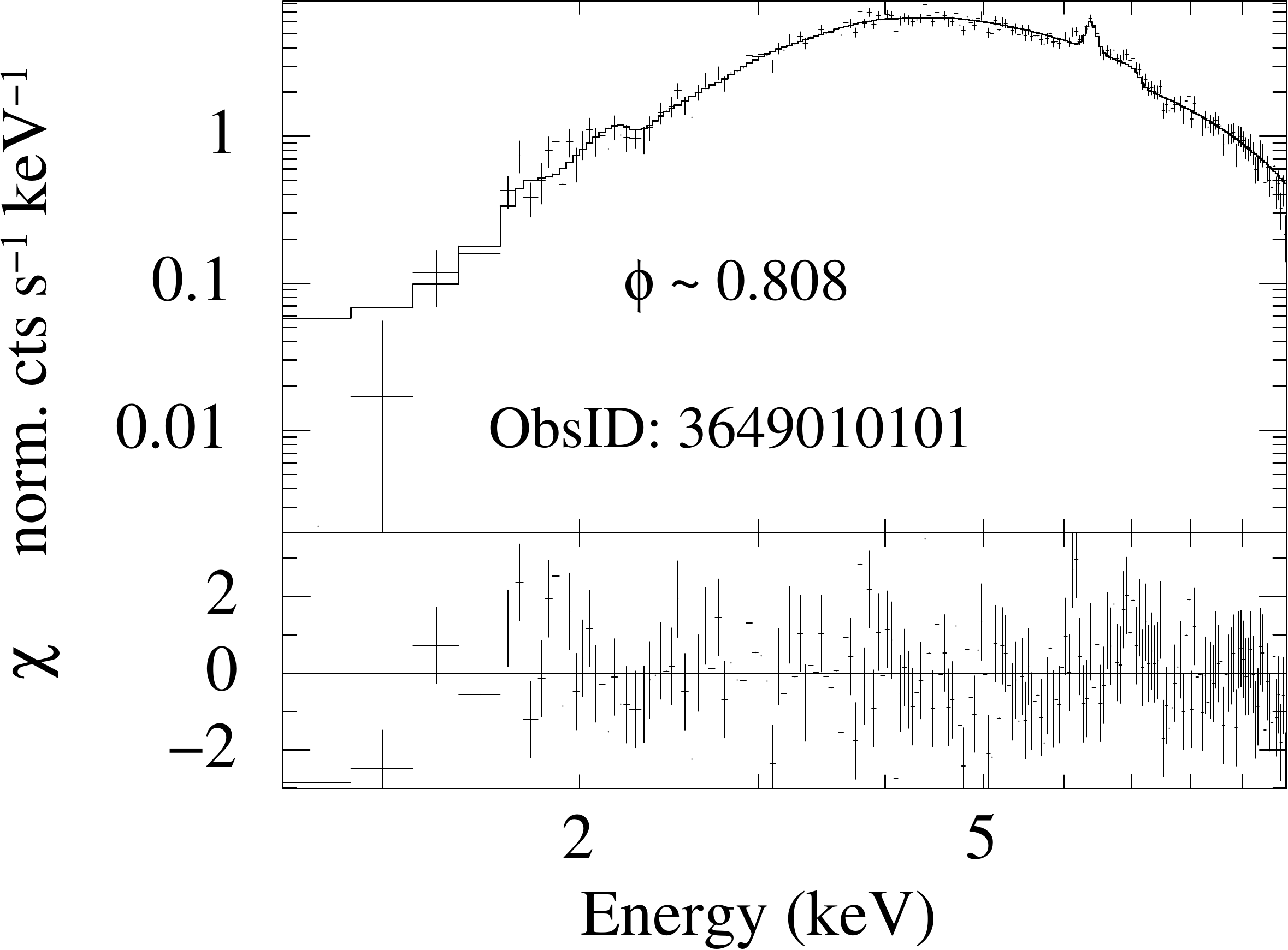} &
\hspace{-0.7cm}
\includegraphics[height=1.5in,width = 1.8in,angle=0]{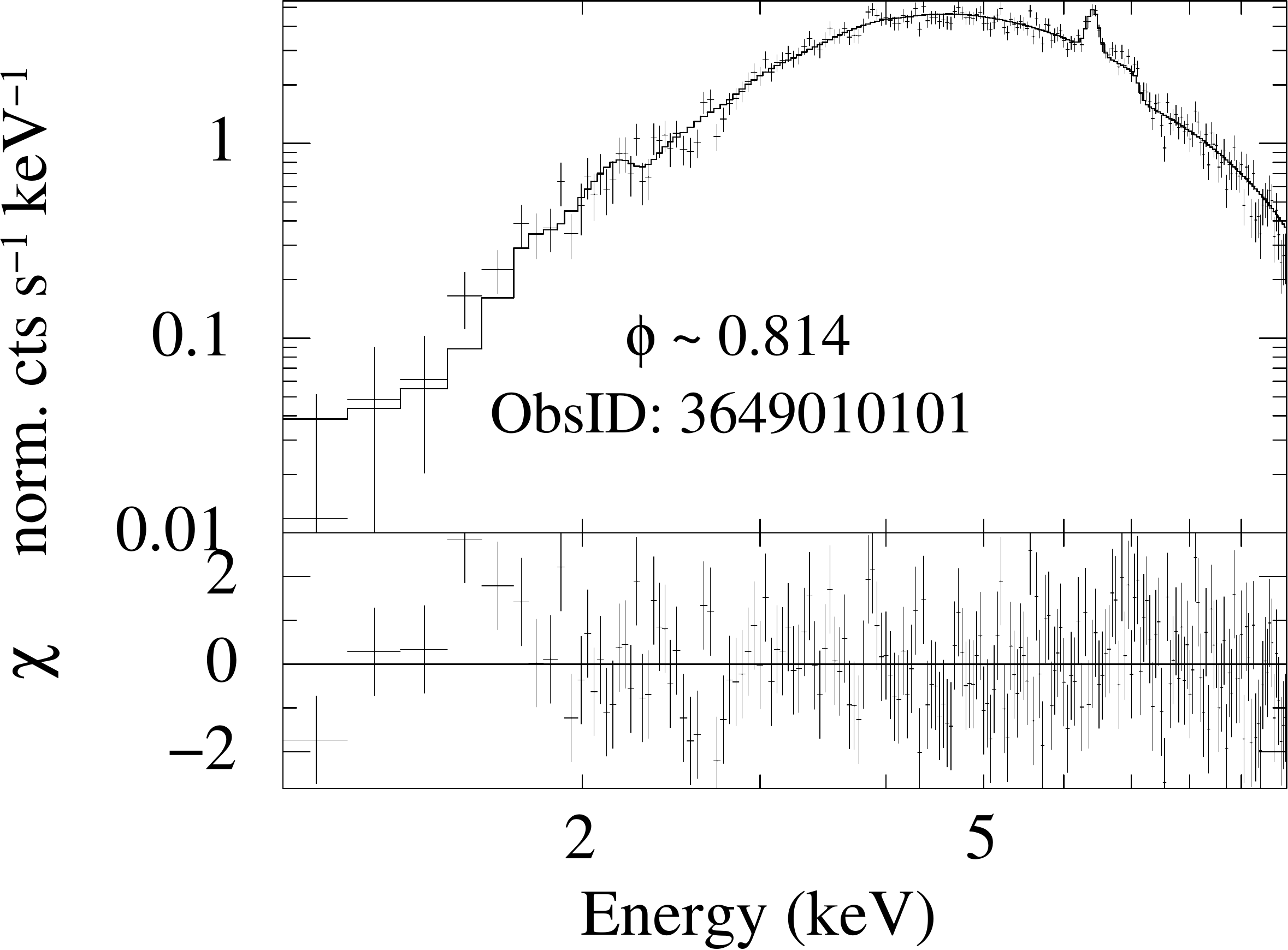} &
\hspace{-0.7cm}
\includegraphics[height=1.5in,width = 1.8in,angle=0]{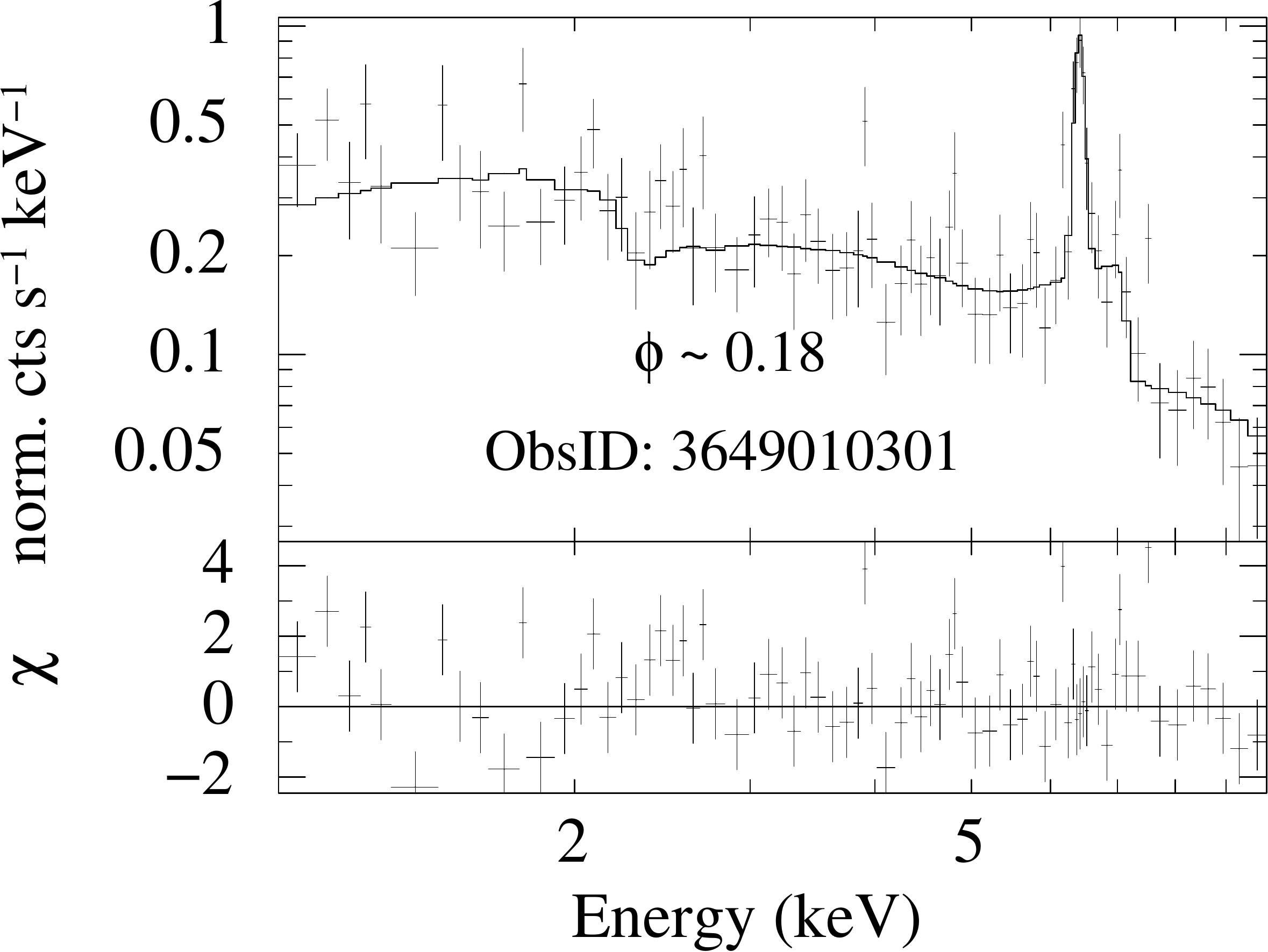}\\
\hspace{-2.0cm}
\includegraphics[height=1.5in,width = 1.8in,angle=0]{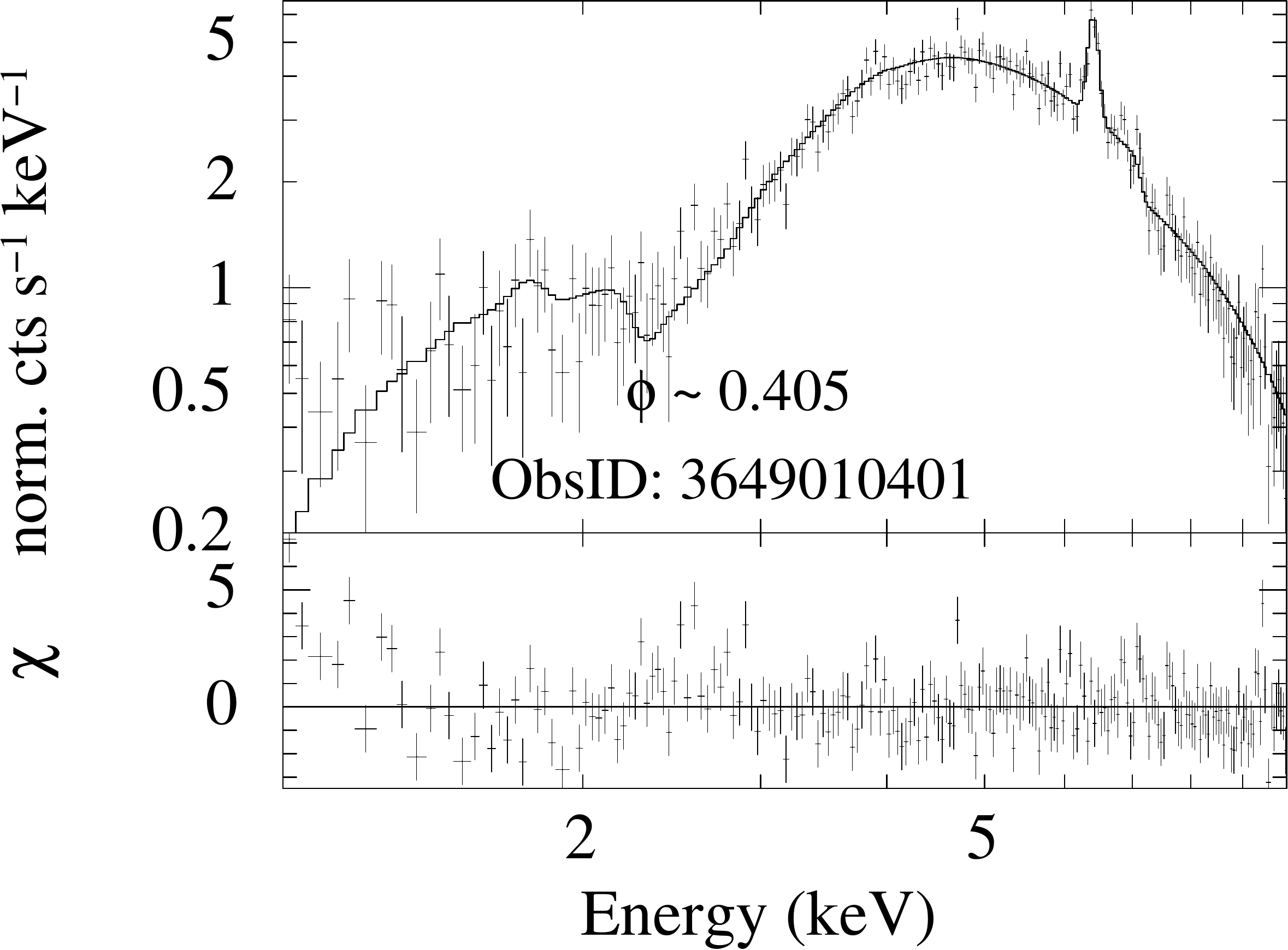} &
\includegraphics[height=1.5in,width = 1.8in,angle=0]{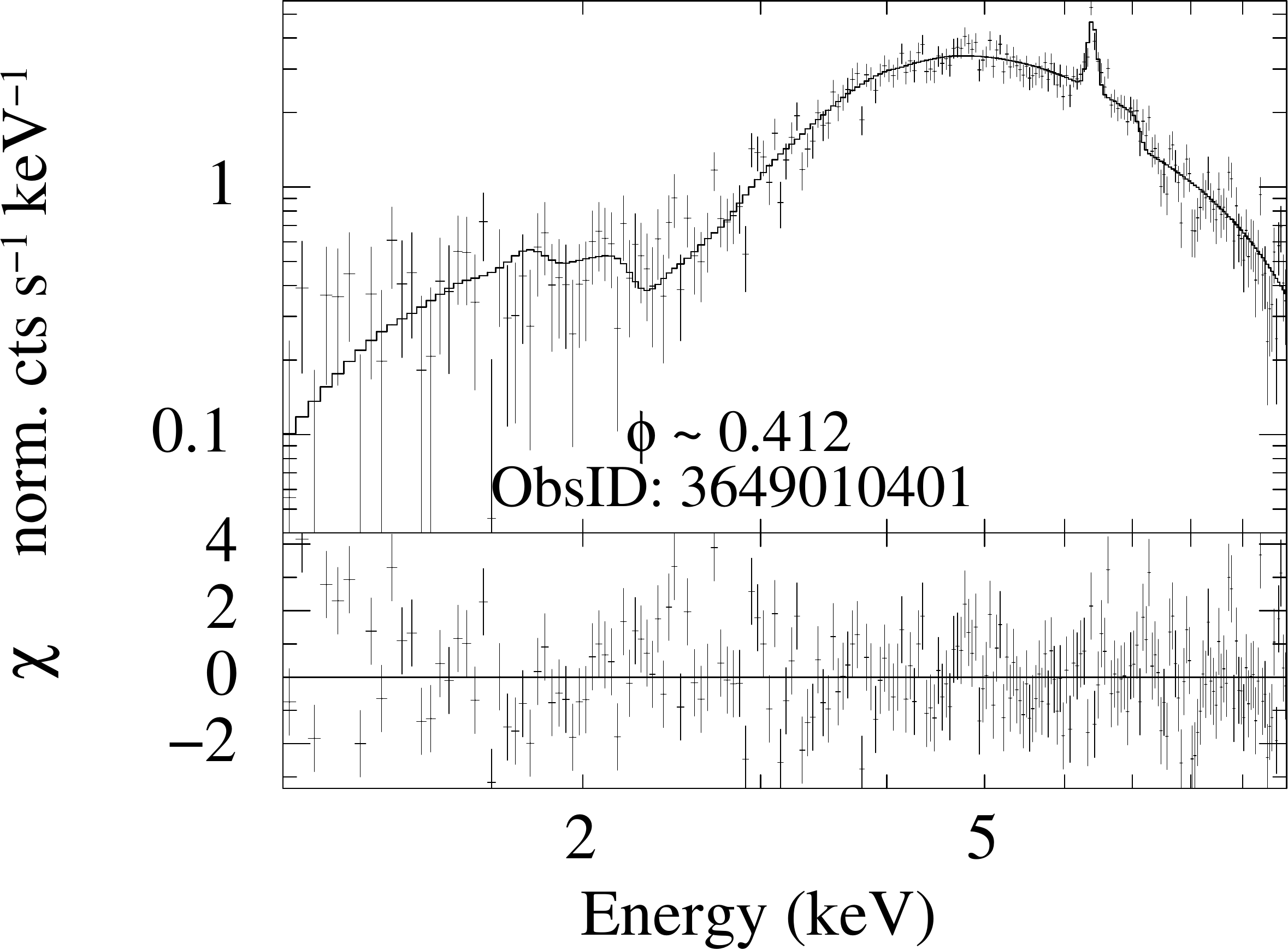} &
\includegraphics[height=1.5in,width = 1.8in,angle=0]{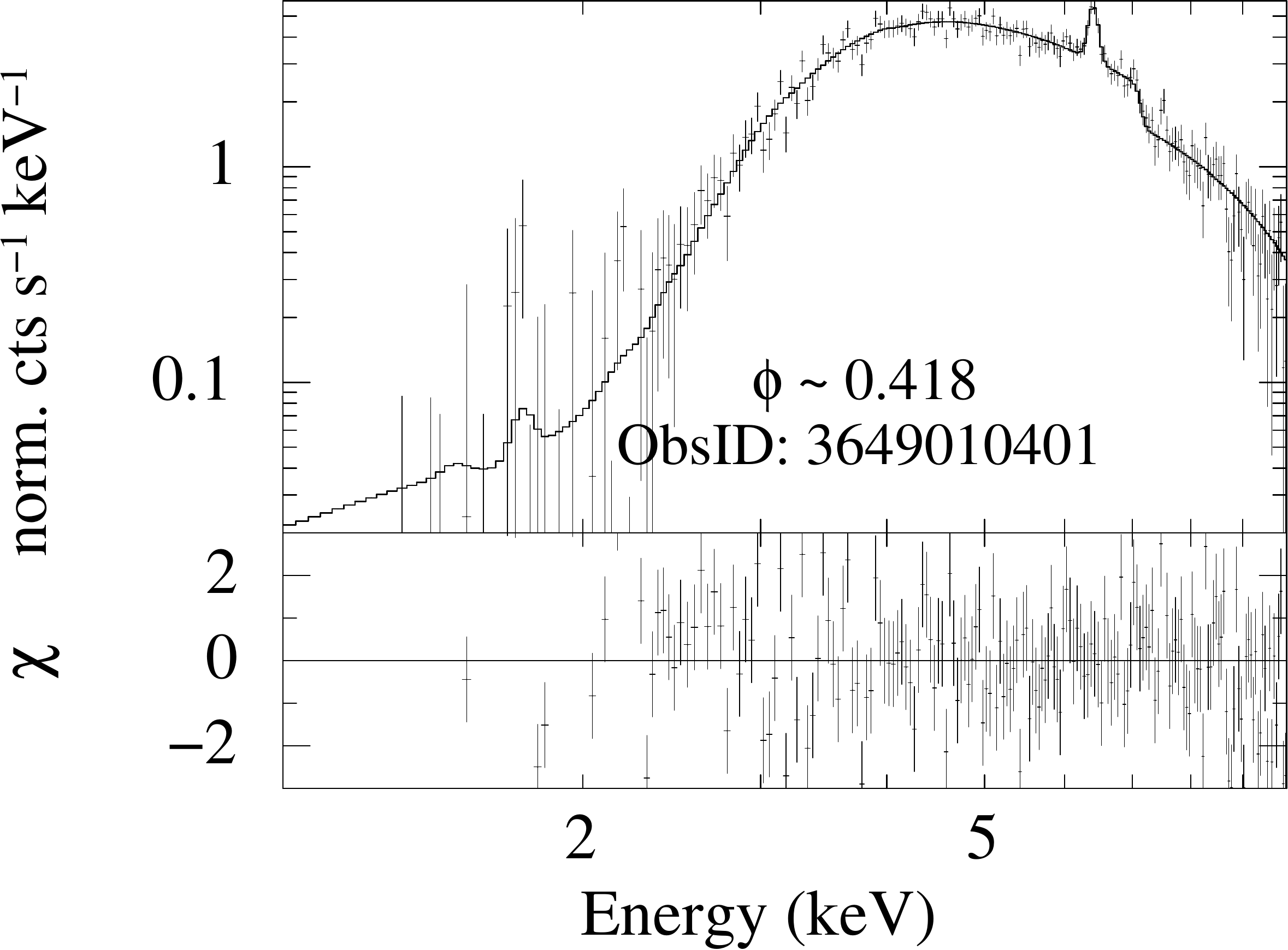} &
\includegraphics[height=1.5in,width = 1.8in,angle=0]{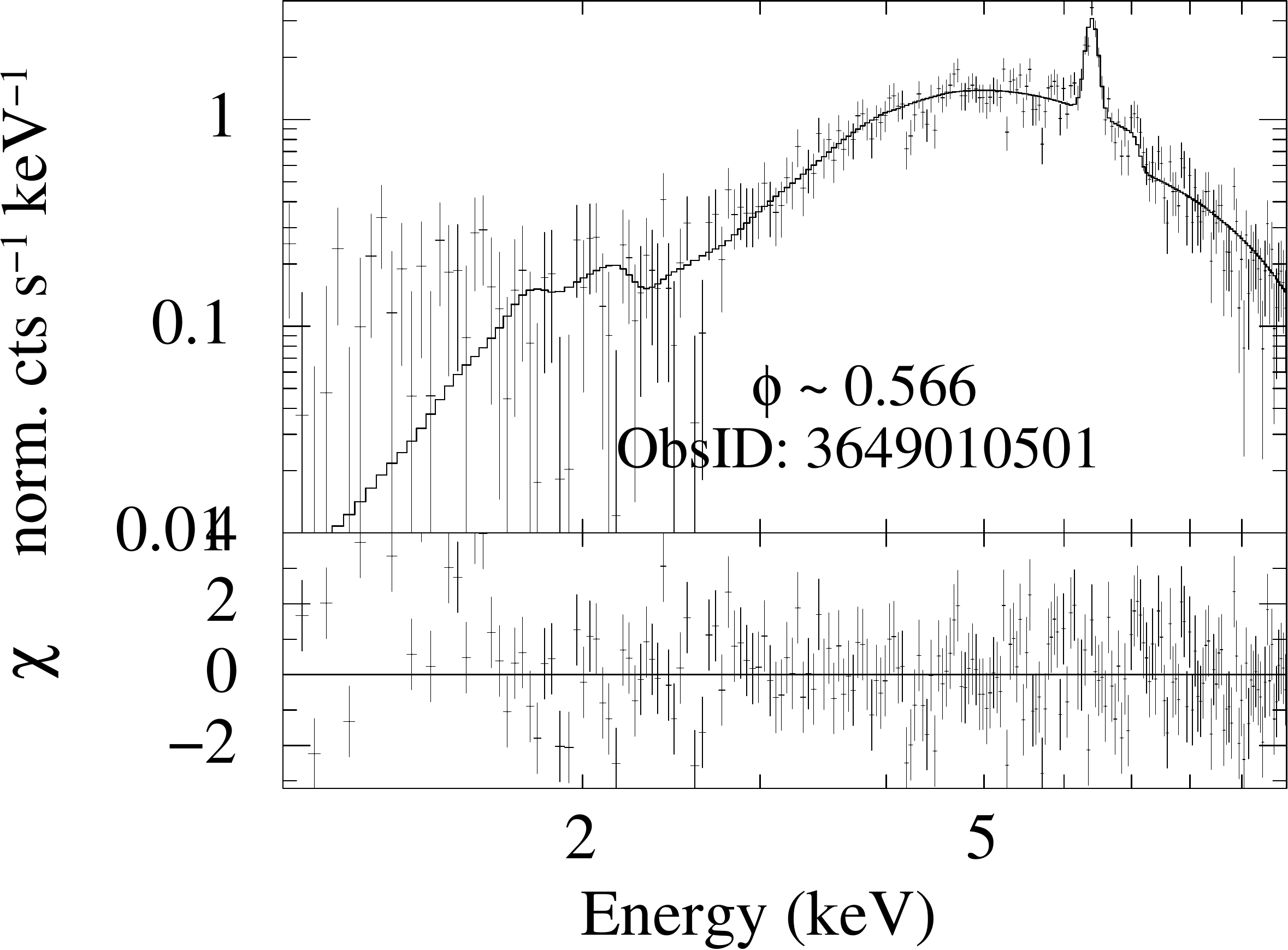}\\
\hspace{-2.0cm}
\includegraphics[height=1.5in,width = 1.8in,angle=0]{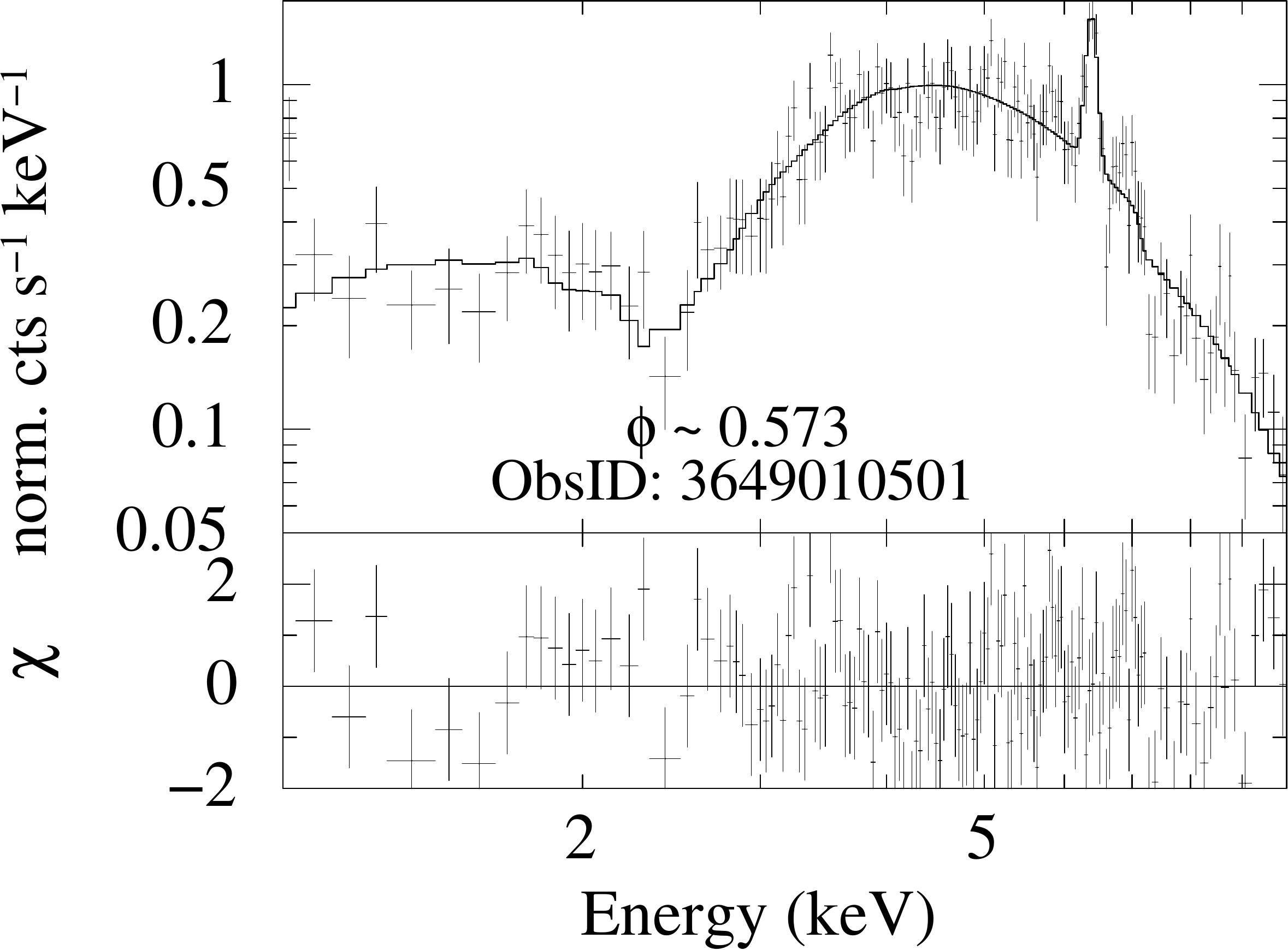} &
\includegraphics[height=1.5in,width = 1.8in,angle=0]{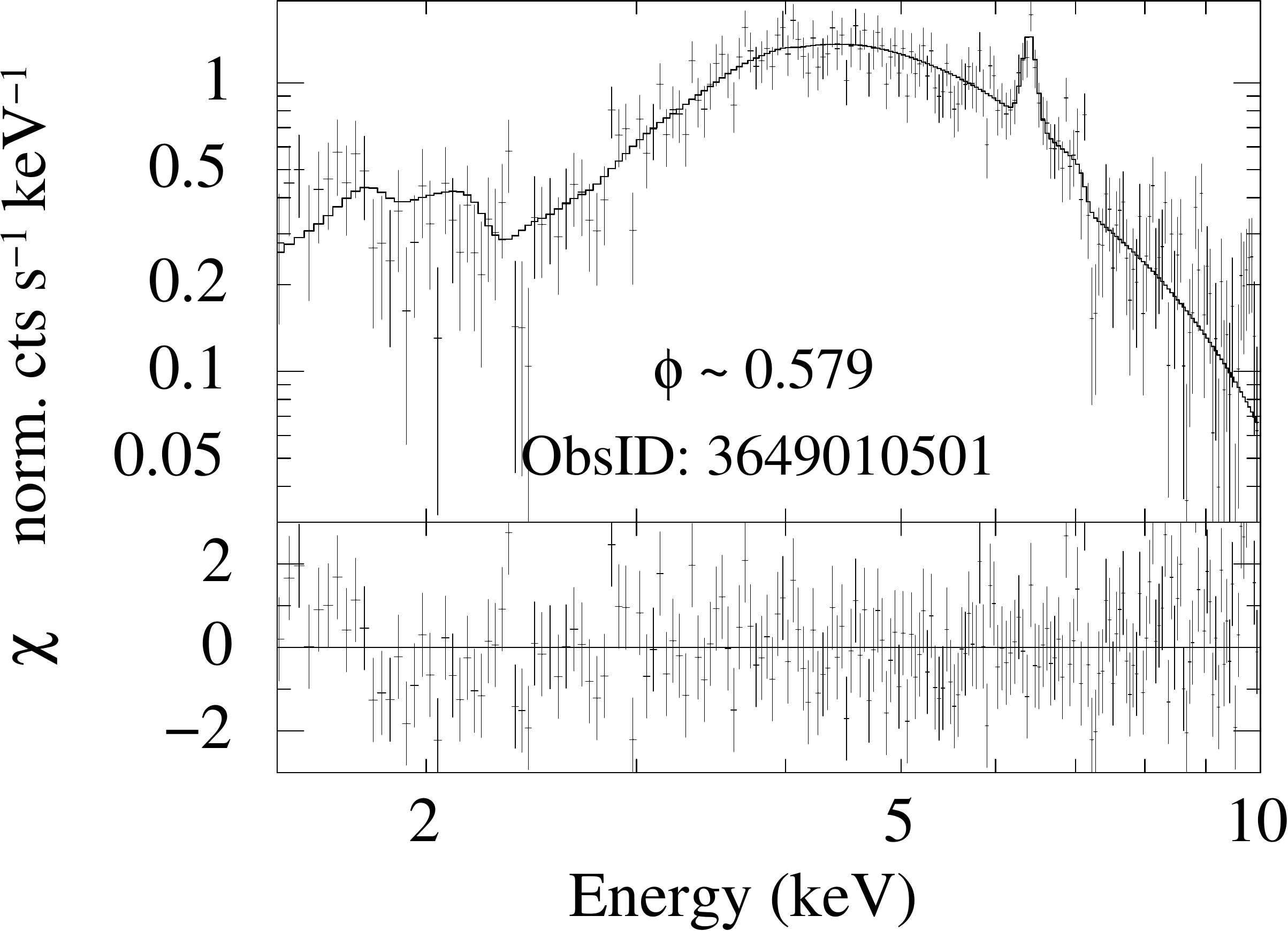} &
\includegraphics[height=1.5in,width = 1.8in,angle=0]{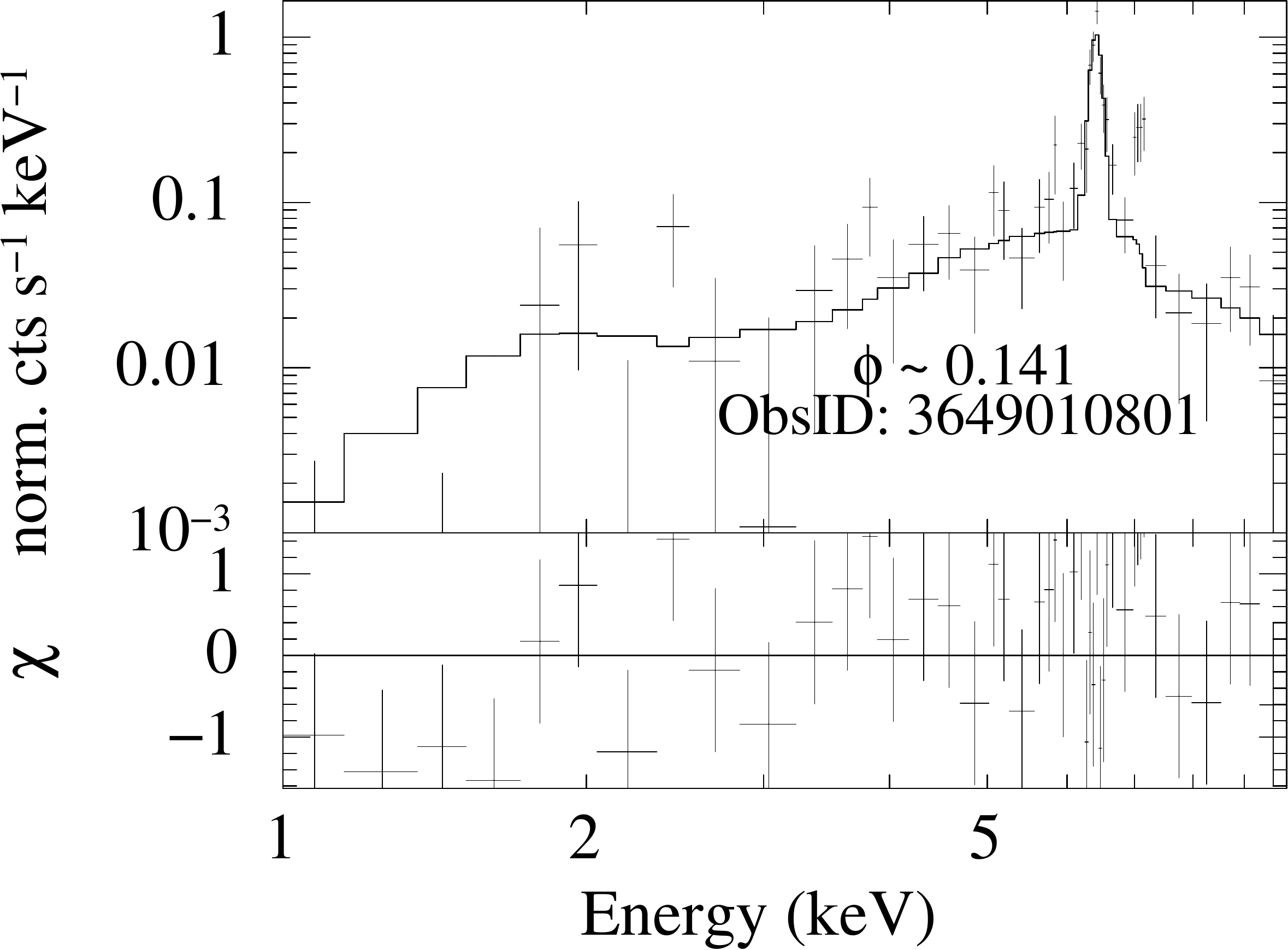} &
\includegraphics[height=1.5in,width = 1.8in,angle=0]{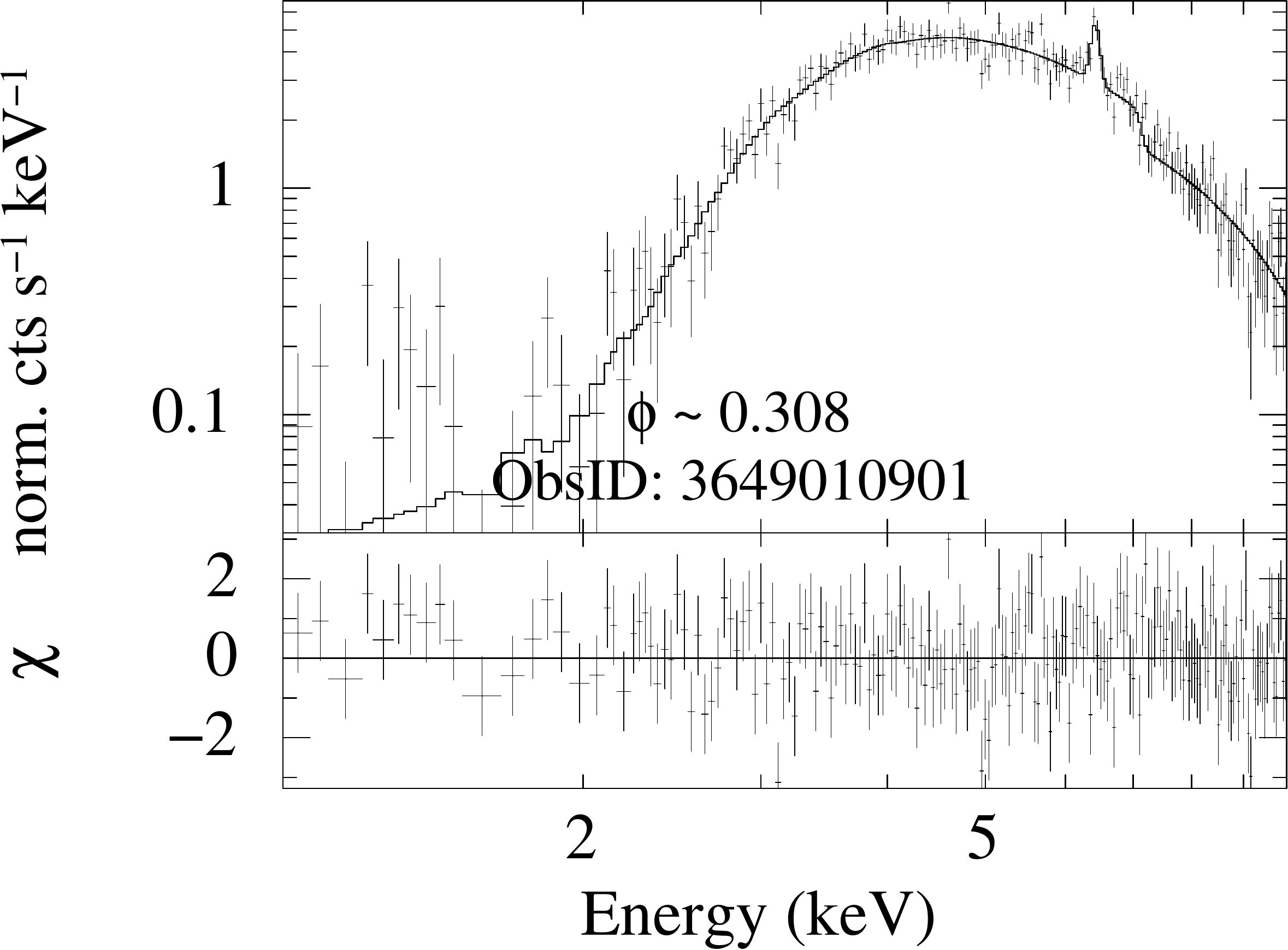}\\
\hspace{-2.0cm}
\includegraphics[height=1.5in,width = 1.8in,angle=0]{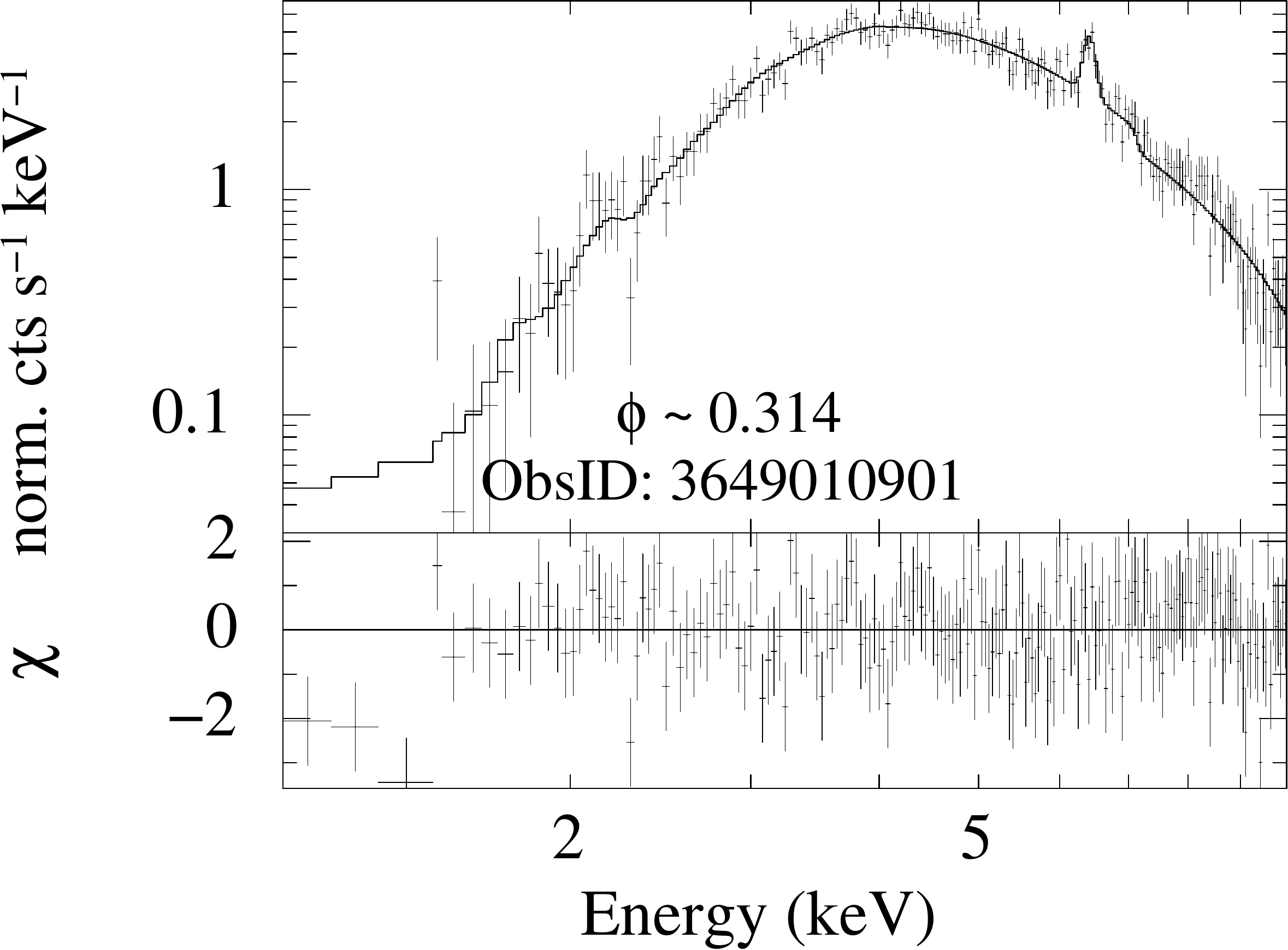} &
\includegraphics[height=1.5in,width = 1.8in,angle=0]{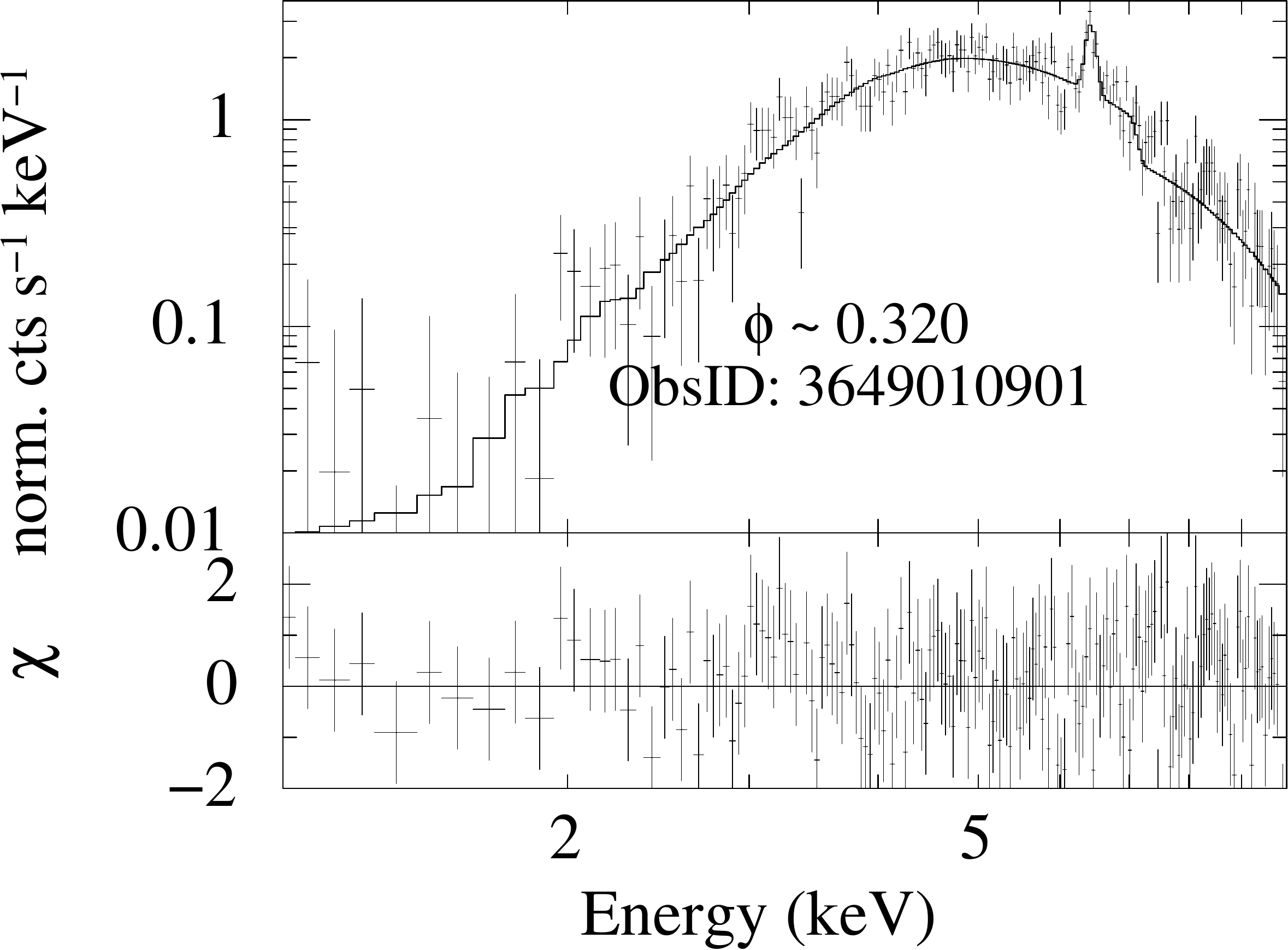} &
\includegraphics[height=1.5in,width = 1.8in,angle=0]{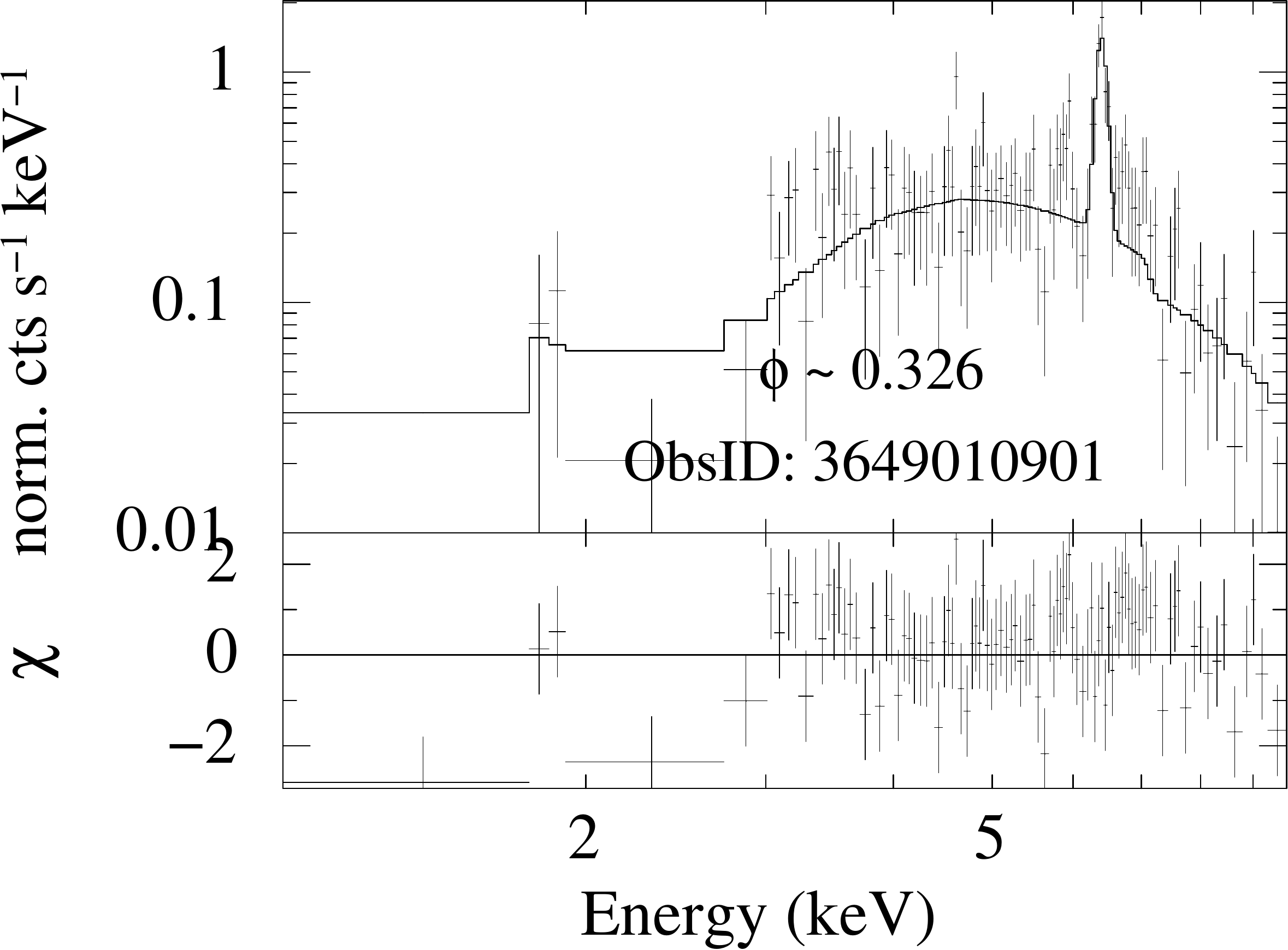} &
\includegraphics[height=1.5in,width = 1.8in,angle=0]{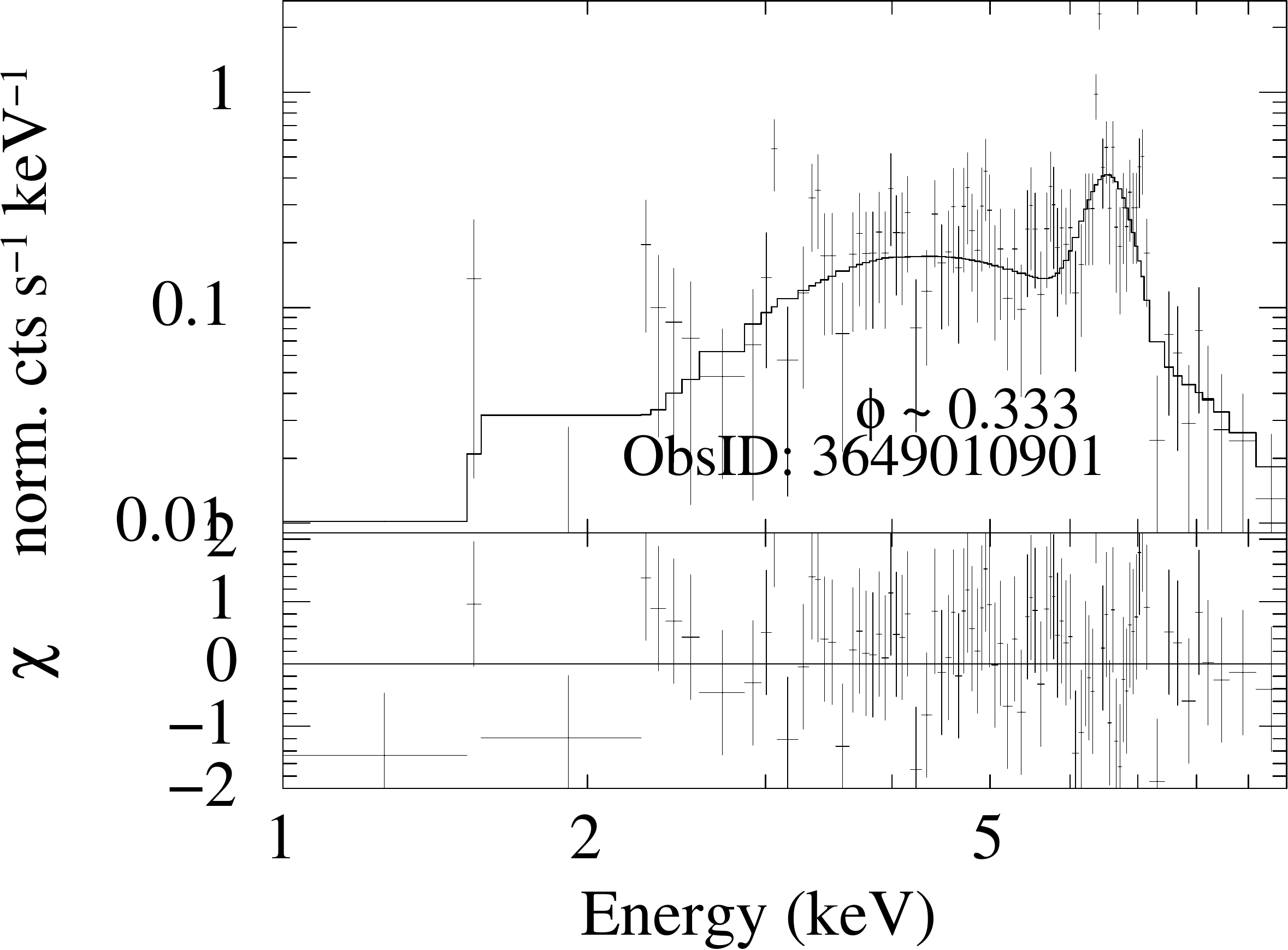}
\end{tabular}
\caption{NICER spectra of \src at different orbital phases, with ObsID and relevant orbital phases noted inside the figures.}
\label{nicer-spectra}
\end{figure}

\begin{figure*}
\centering
\hspace{-1cm}
\includegraphics[height=8cm,width=10cm,angle=0]{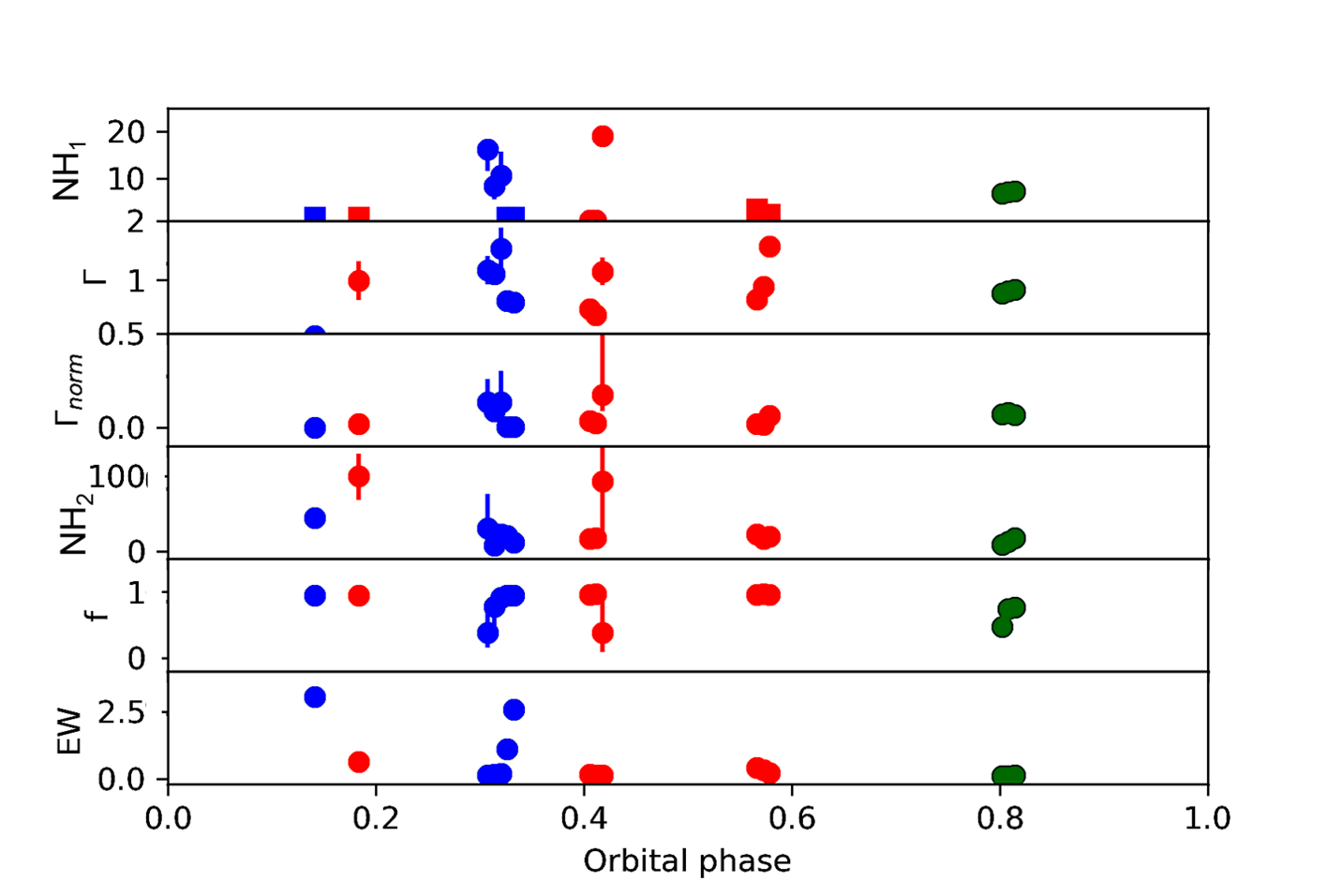}
\caption{Results of \nicer~monitoring of \src over three different orbits of \src~shown by three colors. From top to bottom are global absorption column density (\nho; in units of 10$^{22}$ cm$^{-2}$), spectral index $\Gamma$, normalization of spectral index ($\Gamma_{\rm norm}$; in photons/keV/cm$^2$/s at 1 keV), local absorption column density (\nht; in units of 10$^{22}$ cm$^{-2}$), and equivalent width of iron line (EW; in keV). The square symbols in the uppermost plot indicate the values of \nho~where the errors could not be determined and were frozen to its best-fit values. } 
\label{nicer-plot}
\end{figure*}

We fit the 1-10\,keV spectra from individual observations of NICER with a partial covering absorbed power law for global and local absorption (i.e., \nho~and \nht) along with a Gaussian for the iron K$\alpha$ line.
One ObsID, 3649010801, showed hints of highly ionized lines at 6.97\,keV and 6.7\,keV two more Gaussian were used to fit these lines. The individual spectral plots are shown in Fig.~\ref{nicer-spectra} and the variation of spectral parameters with time and phase are shown in Fig.~\ref{nicer-plot}. 

\begin{figure*}
\flushleft
\hspace{-1.1cm}
\includegraphics[scale=0.7,angle=-90]{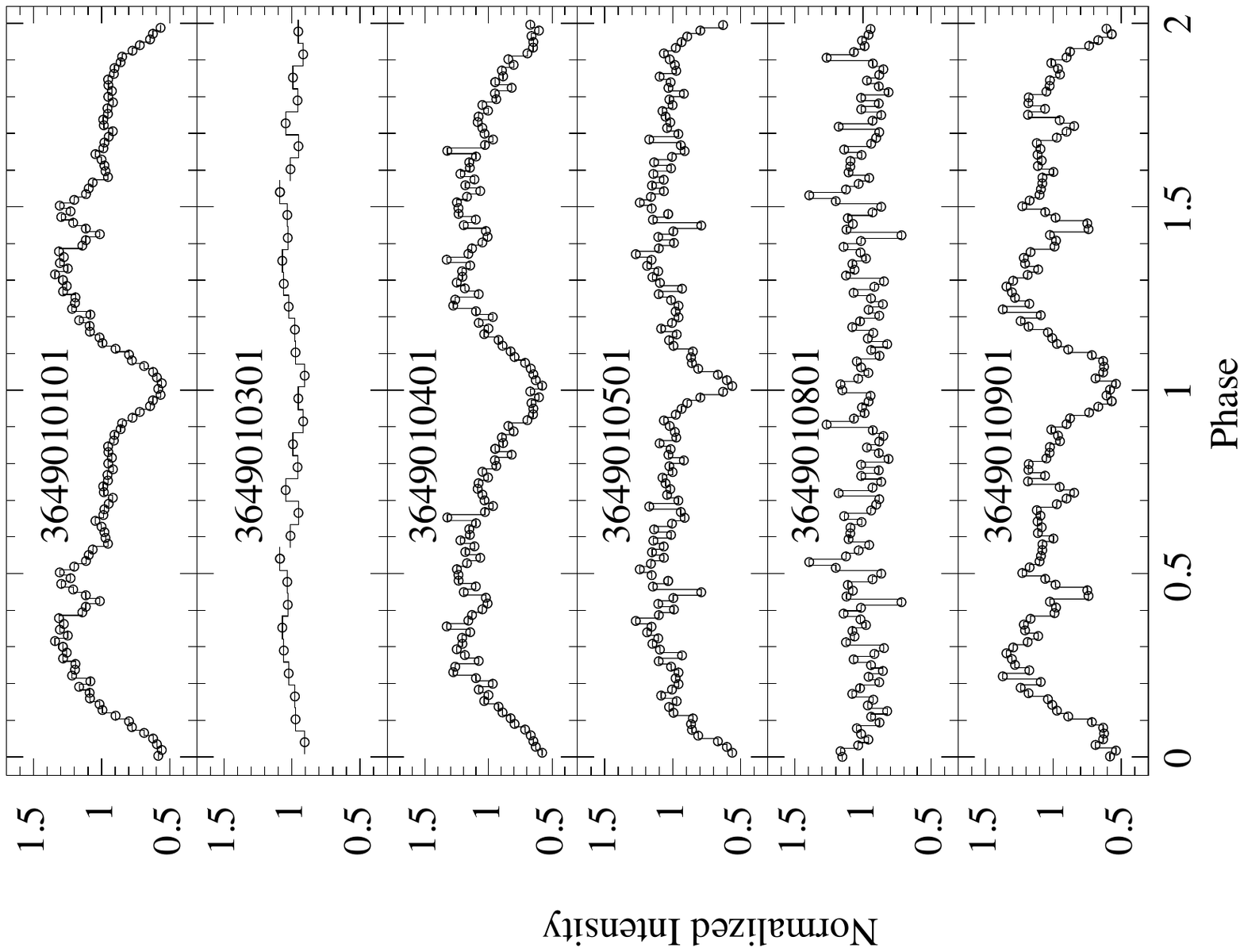}

\vspace{2cm}
\caption{NICER pulse profiles for different ObsIDs indicated in the plot. Pulsations are not detected in highly absorbed segments 3649010301 and 3649010801.}
\label{nicer-pp}
\end{figure*}

The \nicer~monitoring of \src~over its three different orbital cycles of the source (spectra in Fig.~\ref{nicer-spectra}) reveals large variations in the absorption column density with orbital phase, (see Fig.~\ref{nicer-plot}).
We find that the pulsations are present in at least 4 of the 6 ObsIDs (\nicer~pulse profiles in Fig.~\ref{nicer-pp}). 

The four ObsIDs 3649010101, 3649010401, 3649010501, 3649010901 were folded at epochs  MJD 58961.0003005, 58967.0000643,  58969.0002214, 58977.0 with the periods of 37.096000\,s, 37.064932\,s, 37.084932\,s, 37.055759\,s respectively obtained from the \nicer~lightcurves. The choice of these epochs approximately coincided the minimum of the first peak of the pulse profile to be at pulse phase zero.


Although the dip at phase 0.15 seen in \nustar~is not visible, the dip at 0.4 as seen in \nustar~ (this work) and \suzaku~observations \citep{pradhan2014} is clearly visible.

\begin{table}
\centering
     \setlength{\tabcolsep}{3pt}
     \scriptsize
     \caption{Results of the spectral fit in the energy range 3.0-70.0\,keV with the CUTOFFPL model. All errors are quoted for 90\% confidence level. We froze the value of \nho~to the line of sight absorption of 1.8 $\times$ 10$^{22}$ cm$^{-2}$. From left to right, the local absorbing column density \nht, covering fraction \rm{f}, spectral index $\Gamma$, normalization of the power law ($\Gamma_{\rm{norm}}$; units of photons/keV/cm$^2$/s at 1 keV), folding energy, E$_{\rm{fold}}$ followed by energy, width, equivalent width of iron line and the reduced chi-square per degrees of freedom.}
    \begin{tabular}{ccccccccc}
 
    \hline
       $N_\textrm{H2}$  & f &  $\Gamma$ &   $\Gamma_{\rm{norm}}$ & $E_\textrm{fold}$ &  $E_\textrm{Fe}$ & $\sigma_\textrm{Fe}$ & EW   & ${\chi_{r}^{2}}$/d.o.f  \\
 (10$^{22}$ cm$^{-2}$) & (Covering fraction) &  & &(keV) & (keV) &   (keV) & equivalent width (keV) & \\
 
  53 $\pm$ 2 & 0.93 $\pm$ 0.01 & 0.70 $\pm$ 0.04 & 0.020 $\pm$ 0.001 & 18 $\pm$ 0.5 & 6.31 $\pm$ 0.01 & 0.11 $\pm$ 0.02 & 0.26 $\pm$ 0.01 & 1.24/665  \\
      \hline
   \end{tabular}
    
  \label{nustar_spectrum} 
  \end{table}

\section{Discussion}
\label{disc}

\subsection{X-ray variability in \src~caused by clumpy winds}
The light curve of OAO 1657$-$415 as measured by \nustar~display changes in count rate in few\,ks interval seen in many classical SGXBs \citep{walter2015,pradhan2014}. 
The good statistics of \nustar~data allowed us to perform a detailed investigation of the changes in the HR as a function of time. 
In order to better understand the origin of these variations, we first identified 20 time intervals corresponding to the most noticeable HR changes and then used these intervals to perform an HR resolved spectral analysis of the \nustar~data with the same model used for the average spectrum.

In Fig.~\ref{variation-time-res}, it can be appreciated that the rises in the HR (e.g., intervals 2, 5, 17, 18, 19) are largely related to the increases in the local absorption column density (above the baseline value of $\sim$ 50$\times$10$^{22}$ cm$^{-2}$) and increase in the equivalent width of the iron line. At intervals 3-4, 7-8, 12, there is a drop of the absorption below the average value of $\sim$ 50$\times$10$^{22}$ cm$^{-2}$ along with a drop in the HR below its average value of $\sim$ 10. For intervals 17, 18 and 19, there is a dramatic increase in absorption as well as HR. These large variations in absorption column density do not translate into modifications in the photon index and folding energy. It is agreement with previous conclusions in the literature \citep{pradhan2014,jaisawal2014} and leads us to favor the possibility that it is caused by a clumpy stellar wind rather than a varying continuum hardness. Our findings are consistent with the physical scenario in which the soft X-ray photons are intercepted by cold matter (neutral Fe) in the clumpy wind, resulting in an increase in the absorption column density and equivalent width of the Fe line.

During the segments (i.e., 17-19), there is also a decrease of the X-ray count rate compared to its preceding segments (i.e., 13-16). Such behavior where absorption increase during X-ray flux decays has been previously observed in other classical SgXBs like IGR J18027-2016 \citep{pradhan2019_igrj18027} as well as in most Supergiant Fast X-ray Transients \citep{bozzo2016,bozzo2017_gate}. The proposed interpretation is also related to the clumpiness of the stellar wind. The increase in the absorption~close to the rise of the X-ray flux can be an indication that the event is being triggered by a dense clump partly intercepting the orbit of the compact object along the line of sight to the observer. When the accretion of the clump material begins, the X-ray flux increases and causes a partial photoionization of the clump surrounding the neutron star. This leads to a rapid decrease of the absorption, which returns to values measured before the flux only when the X-ray fluxes decrease along the decay of the flare (the value of absorption~after the flare depends also on the location of the re-arranging of the accreted clump with respect to the line of sight to the observer). Such an interpretation has been proposed for \src~earlier in the literature, which we discuss as follows. During the \suzaku~observation spanning orbital phase of 0.1-0.3, the absorption varied in the range $\sim$ 10-100 $\times$ 10$^{22}$ cm$^{-2}$ and the correspondingly measured equivalent width of Fe K$\alpha$ line was $\sim$ 0.1-1.0 \,keV \citep{pradhan2014}. During an \chandra~observation carried out in phase 0.5-0.6, the absorption was $\sim$ 70 $\times$ 10$^{22}$ cm$^{-2}$ and the corresponding equivalent width  was $\sim$ 1\,keV \citep{oao_hetg2019}. In the current observation, although the average equivalent width  is lower than previous observations, we find that the equivalent width  of the iron line scales with the absorption as expected. Such a variation is fully consistent with fluorescence emission of X-rays in the clumps of matter around the neutron star (e.g., Fig.~1 of \citealt{pradhan2018}). Calculating the 3-70\,keV flux in each HR segment and assuming a distance of 7.1\,kpc for the source \citep{A06}, we find that the X-ray luminosity varies from  L$_X$ $\sim$ 4--10 $\times$ 10$^{36}$ erg~s${-1}$ through the \nustar~observation.

Following the results of pulse phase resolved spectroscopy of \nustar~data (Fig.~\ref{spec-phase}), we argue that the dip in pulse profiles at phase 0.4 corresponds to an increase in \nht~and is therefore possibly caused by obscuration of X-rays. On the other hand, the dip at phase 0.15 could have a different explanation  (see section \ref{dip} for details on this observation). From the spin phase resolved spectroscopy, we also find that the iron line also showed a flux variation with pulse phase, indicating anisotropic matter distribution around the neutron star as expected from clumpy winds. 
Additionally, from \nicer~observations, we note a correlation between absorption and equivalent width  of iron line, which is also consistent with what is seen from earlier measurements in the literature. This scenario is agreeable with reprocessing of X-ray photons in the cold fluorescence matter, possibly clumps in the wind. We will revisit this discussion in sections \ref{sec:kalpha} and \ref{sec:clumpsize} below.

\subsection{Dip at spin phase 0.15: Changes in viewing geometry or a transient accretion disk?}
\label{dip}
We report, for the first time, a dip at spin phase of $\sim$ 0.15 seen in the \nustar~pulse profiles. The dip in \src~lasts for roughly 0.06 of the spin phase (i.e., $\sim$ 2\,s) and, albeit variable in different time segments (see, Fig.~\ref{pp}), is visible throughout the \nustar~observation of $\sim$ 150\,ks. The strength of the dip is energy-dependent, being the most prominent between 5-30\,keV (see, Fig.~\ref{energy-pp}), the same energy range in which the X-ray photons undergo inverse Compton scattering and produce the high energy cut-off in the X-ray spectrum of HMXBs. Although the occurrence of dips in pulse profiles are not unusual when the X-rays are intercepted by clumps in the stellar wind, such dips caused by obfuscation in clumps are short-lived and are often accompanied by a large increase in absorption (e.g., dip at phase 0.4 in Fig.~\ref{spec-phase}). From the spin phase resolved spectroscopy, we see that there is no increase in absorption around phase 0.15, but rather this dip is accompanied by a change in folding energy at the same phase (see, Fig.~\ref{spec-phase}). These findings, coupled with the fact that the dip last throughout the entire \nustar~observation of $\sim$ 150\,ks exposure, the likely explanation is that the dip at 0.15 is caused by a change in viewing geometry — possibly X-ray emission in multiple accretion columns or spectral changes due to the formation of a transient disk.

Note that in earlier analysis of the same \nustar~data (\citealt{saavedra2022,sharma2022}), the period at which the light curves are folded are determined from individual time segments. In our case, we have performed the orbital correction followed by determining the spin period from the whole data set, which is then used to fold the light curves. The advantage of the latter approach (versus search over smaller time segments) is that the spin period is determined with greater accuracy. This will therefore allow us to study the evolution of pulse profile morphology over the duration of the observation more accurately. We also confirmed the spin period derived here in this work, after correcting for the orbital motion, is consistent with the Fermi/GBM measurement interpolated to the time of this observation.

This dip in \src~is also reminiscent of the dip discovered in the \xmm~observations of the Be X-ray binary, EXO 2030+375 during the Type-I outburst caused by self-obscuration of the accretion stream passing through the observer line of sight  \citep{Ferrigno_2016}. While a similar possibility of an accretion stream passing the observer line of sight in \src~cannot be ruled out, given that the dip is energy dependent and most prominent in the critical energy range of 5-30\,keV in which the photons undergo inverse Comptonization to produce a cut-off in the X-ray spectrum, the former possibility of spectral changes due to X-ray emission in multiple accretion columns or the formation of a transient accretion disk are more likely as argued earlier.

We also discovered from the Fermi/GBM monitoring\footnote{https://gammaray.nsstc.nasa.gov/gbm/science/pulsars/lightcurves/oao1657.html.} that during the time of the \nustar~observation the pulsar was spinning down for $\sim$ 300 days after a long spin-up of $\sim$ 800 days. Since \src~is presumed to host a transient accretion disk, we cannot exclude the possibility that the spin up followed by the spin down of the pulsar (as seen from GBM monitoring) could be related to changes in the emitting region caused by a transient accretion disk. The signatures of a transient accretion disk are manifested as changes in the X-ray spectrum and the pulse profiles has earlier been seen for GX 301-2 with \nustar~\citep{Nabizadeh2019}. Additionally, the very low terminal wind speed in \src~strongly favors the formation of wind-captured disks due to the significant shearing of the flow \citep{shapiro1976}. In \citet{elmellah2019}, the authors argue that transient disks are likely to form when the stellar wind speed at the orbital separation is of the order of or lower than the orbital speed, which is the case in \src. \citet{taani2019} also singled out \src~as a HMXB susceptible to host a wind-captured disk, given its orbital and wind parameters. A better understanding of accretion-induced torques is required to understand whether the torque reversal captured by GBM could have been provoked by a transient wind-captured disk.

As argued earlier, while absorption of soft X-rays in clumps could cause dips in the pulse profiles, in such cases, one would also expect the pulse profiles to change with hardness ratio, the same is not seen here (see Fig~\ref{pp}, Fig~\ref{pp-heatmap}). This is probably due to the fact that the \nustar~band pass is dominated by hard X-rays, where absorption has lesser energy dependence. We refer the readers to Fig.~2 of \citet{hirsch2019} for the effect of absorption on the soft X-ray spectral shape where a partial covering remove flux at soft energies making only the direct component visible.

Earlier results have revealed that the lack of pulsations (pulse fraction less than 2\%) in the soft X-ray emission of \src across the \chandra~observation \citep{oao_hetg2019} and during the time interval of the \suzaku~observation with the highest absorption ($\sim$ 10$^{24}$ cm$^{-2}$ in interval `C'; \citealt{pradhan2014}) can be ascribed to the effect of dense clumps obscuring the X-ray source. Compared to these data, the \nustar~observation shows a much lower averaged absorption ($\sim$ 5$\times$10$^{23}$ cm$^{-2}$)  and it is indeed possible that during the \nustar~observation the neutron star was passing through a less dense region of the stellar wind, as expected for any reasonable assumption of the properties of the wind of the O supergiant (see e.g., \citealt{nunez17} and references therein). The low density of the accretion environment during the \nustar~observation explains the fact that in none of the 20 HR resolved time intervals, we observed a disappearance of pulsations (see Fig.~\ref{pp} for pulse profiles of the segments). On the other hand, the \nicer~observations reveal that pulsations were detected in 4 of the 6 ObsIDs. The two ObsIDs (3649010301 and 3649010801) are highly absorbed ($\sim$ 10$^{24}$ cm$^{-2}$) and no pulsations were detected during these observations. 

Note that a relatively recent class of sources, named the ultraluminous X-ray pulsars (ULXPs), exhibit strong variability in the pulsed fractions. Such pulse dropout phenomena have been observed in some ULX pulsars: M82 X-2 \citep{Bachetti2014}, NGC 7793 P13 \citep{furst2016,israel2017a}, NGC 5907 ULX-1 \citep{israel2017b}, and NGC 300 ULX1 \citep{carpano2018}. While the most plausible explanation for such a behavior has been proposed to be the onset of the propeller effect \citep{tsygankov2016}, the exact mechanism for pulse dropout is still not established. Interestingly, such pulse dropouts uncorrelated with flux have been reported for some HMXBs as well: Vela X-1 \citep{Kretschmar2000AIPC}, LMC X-4 \citep{Brumback_2018}, SMC X-1 \citep{pike2019, Pradhan2020} and \src\citep{jaisawal2021}. For all the four sources, the pulse-dropout is unlikely to be caused by the onset of the propeller regime, since the cessation of pulsations were not caused by cessation of accretion/X-ray flux variations. Note also that the pulse dropouts in all these three sources occur in hard X-rays and are therefore unlikely to be caused by increased absorption in an intervening matter (e.g., clumps).

\subsection{Where are the Iron lines formed in \src? }
\label{sec:kalpha}
Given the high inclination of the source, the compact object in \src~undergoes eclipses for $\sim$ 20\% of the orbital period. Using ASCA observations, \cite{A06} made the first estimate for Fe K$\alpha$ line-forming region as 19 ($\pm$ 12) lt-sec from the compact object. This value was based on the intensity variations of Fe K$\alpha$ line in and out of eclipse. Similar results were also obtained under the assumption that clumps are composed of neutral hydrogen in a moderately ionized surrounding \citep{jaisawal2014}, limited by ionization parameter \citep{kallman1982} below. 
\begin{equation}
  \xi= \frac{L_X} {nR^2} \leqslant 10^{2} {\rm ~ in~erg~cm/s}
  \label{xi}
\end{equation}
  where $L_X$ is the X-ray luminosity of the source, $n$ is the free electron density in the clump at a distance $R$ from the compact object. It should however be noted that the luminosity of the source is variable and for the \chandra~observation at orbital phase of $\sim$ 0.55, where the luminosity was two orders of magnitude less than the \suzaku~observation, the limiting inner radius of the Fe K$\alpha$ line-forming region was found to be much closer to the neutron star at $\sim$ 2.5 lt-sec \citep{oao_hetg2019}. Note however that the \chandra~observation was carried out in a `special' phase where long-term \integral~light curves show a dip when the compact object is completely enshrouded by dense (and large, possibly phase-locked) matter \citep{BS08,oao_hetg2019}.

Using our current measurements, we now try to constrain the regions of neutral and ionized iron line formation in \src. Assuming that the inter-clump medium is void, the volume filling factor $f_{vol}$ can be used to determine the number density $n$ from $n=n_{\rm wind}/f_{\rm vol}$ \citep{sundqvist2018}, where $n_{\rm wind}$ is the stellar wind density at the orbital separation if the wind were smooth and isotropic, given by: 


\begin{equation} 
n_{\rm wind} =	\frac{\dot{M}/m_{\rm p}}{4 \pi a^2 v_a},
\end{equation} 

\noindent
where $\dot{M}$ is the mass-loss rate, $m_{\rm p}$ is the mass of the proton, $a$ is the orbital separation, and $v_a$ is the wind velocity at a distance $a$ from the donor star. Working along the line joining the star to the neutron star, we can rewrite Eqn.\,\eqref{xi}:

\begin{equation}
  \xi= \xi _0\left(\frac{R}{{a-R}}\right)^2,
  \label{eq:r100}
\end{equation}

\noindent
where:

\begin{equation}
    \xi _0=4\pi f_{\rm vol} \frac{m_p v_a L_X}{\dot{M}}.
\end{equation}

We look for the distance $R=R_{100}$ to the neutron star where $\xi=100$ erg cm$/$s using Eqn.\,\eqref{eq:r100}. We take a terminal wind speed of 250\,km s$^{-1}$, a stellar radius of 25R$_{\odot}$ and a mass loss rate of 1$\times10^{-6}$ M$_{\odot}$ yr$^{-1}$ \citep{mason2012}. We find that $R_{100}$ ranges from 20 to 40 lt-s for nominal volume filling factors, f$_{\rm vol}$ = 0.1-0.3 \citep{sundqvist2018}, and~for~L$_X$ = 4-10 $\times$ 10$^{36}$ erg~s${-1}$. The luminosity is calculated from the 3-70\,keV fluxes in HR segments for a distance of 7.1\,kpc \citep{A06}. This value of line-forming region of K$\alpha$ line is consistent with the value obtained by \cite{A06}. However, it is to be understood as a lower limit since we worked along the line joining the two bodies but at the orbital phase of the current \nustar~observation (0.612-0.743), it does not correspond to the line-of-sight. If the geometry matches the classic one considered by \cite{Hatchett77}, the latter intercepts the $\xi=100$ erg cm$/$s surface at a distance from the neutron star larger than $R_{100}$.

Among the \nicer~observations (see, Fig.~\ref{nicer-spectra}), one ObsID 3649010801 showed hints of ionized lines, and we fit this spectrum to obtain the ratio of the 6.97 to the 6.7\,keV line $\sim$ 3.  This translates to $\xi \ge 10^{3.8}$ (refer to Figure 8 of \citealt{ebisawa1996}). Thereafter, using Eqn.~\ref{xi}, the line emitting region $R$ for these ionized lines for $n$ $\sim$ 3$\times$$10^{13}$ cm$^{-3}$ is $\sim$ 7$\times$10$^8$ \,cm ($\sim$ 0.02 lt-s). The ionized lines form much closer to the compact object than the neutral Fe K$\alpha$ lines.

\subsection{Non-detection of a CRSF line}\label{sec:no-cyclo}
We found that the phase-averaged \nustar~spectrum of \src~can be described by a relatively hard power-law continuum modified by an exponential cutoff plus a partial covering component and fluorescence iron K$\alpha$ line at 6.4\,keV. There is no evidence for the previously reported CRSF at 36\,keV (see Section \ref{intro} for details on previous works). We looked for Gaussian like absorption lines in the phase-averaged \nustar~spectrum between 10\,keV and 50\,keV with width of 0.1 times the energy bin and find an upper limit on the optical depth of is about 0.06 (at 90\% confidence level). This value is far smaller than the typical value that is encountered in all other sources with confirmed CRSF \citep{maitra2017,staubert2019,pradhan2021}.

Using the same \nustar~data, \citet{saavedra2022} claimed the detection of a CRSF at
{$\sim$35\,keV with a width of $\sim$29\,keV in the averaged spectrum of the source.} 
The authors however do not make use of a partial covering absorption component as a part of their continuum model, which give rise to wavy structures in the residuals, interpreted as a cyclotron line. We additionally tested the above scenario by fitting the spectrum of \src~ with the continuum parameters frozen to the values reported in \citet{saavedra2022}. We notice an extremely broad and shallow feature extending from 20\,keV up to the highest energy range used for the spectral fit. We are unable to constrain the width of this feature (line centroid fits at $\sim$56\,keV) and this typically represents an inadequacy in the spectral model or incorrect continuum modelling.
While \citet{sharma2022} were able to model a CRSF varying in the range $\sim$ 38--42\,keV with a width $\sim$ 0.8--5\,keV in some time segments, there were no discussions regarding the statistical significance of the feature in the paper except the improvement in $\chi^{2}$ for the spectral fit by adding the multiplicative model `gabs' or `cyclabs'. This is not  adequate to test the presence of the cyclotron line \cite[see][]{O99} and needs to be treated with caution. Note that based on the spin phase-resolved spectroscopy, we see a dip-like feature around spin phase 0.2 at 40\,keV which was reminiscent of the CRSF line, but it is not seen in other phase intervals and is possibly spurious since the F-test reveals the probability of chance improvement (PCI) $\sim$ 0.75. Therefore, in the event of the claimed cyclotron line being model-dependent, the absorption like residuals seen in the X-ray spectrum is likely a result of imperfect continuum modelling - possibly a result of steepening of the spectrum above 20\,keV as discussed in \citet{O99} - and not a true CRSF. 

\begin{figure*}
\centering
\includegraphics[height=9cm,width=10cm,angle=0]{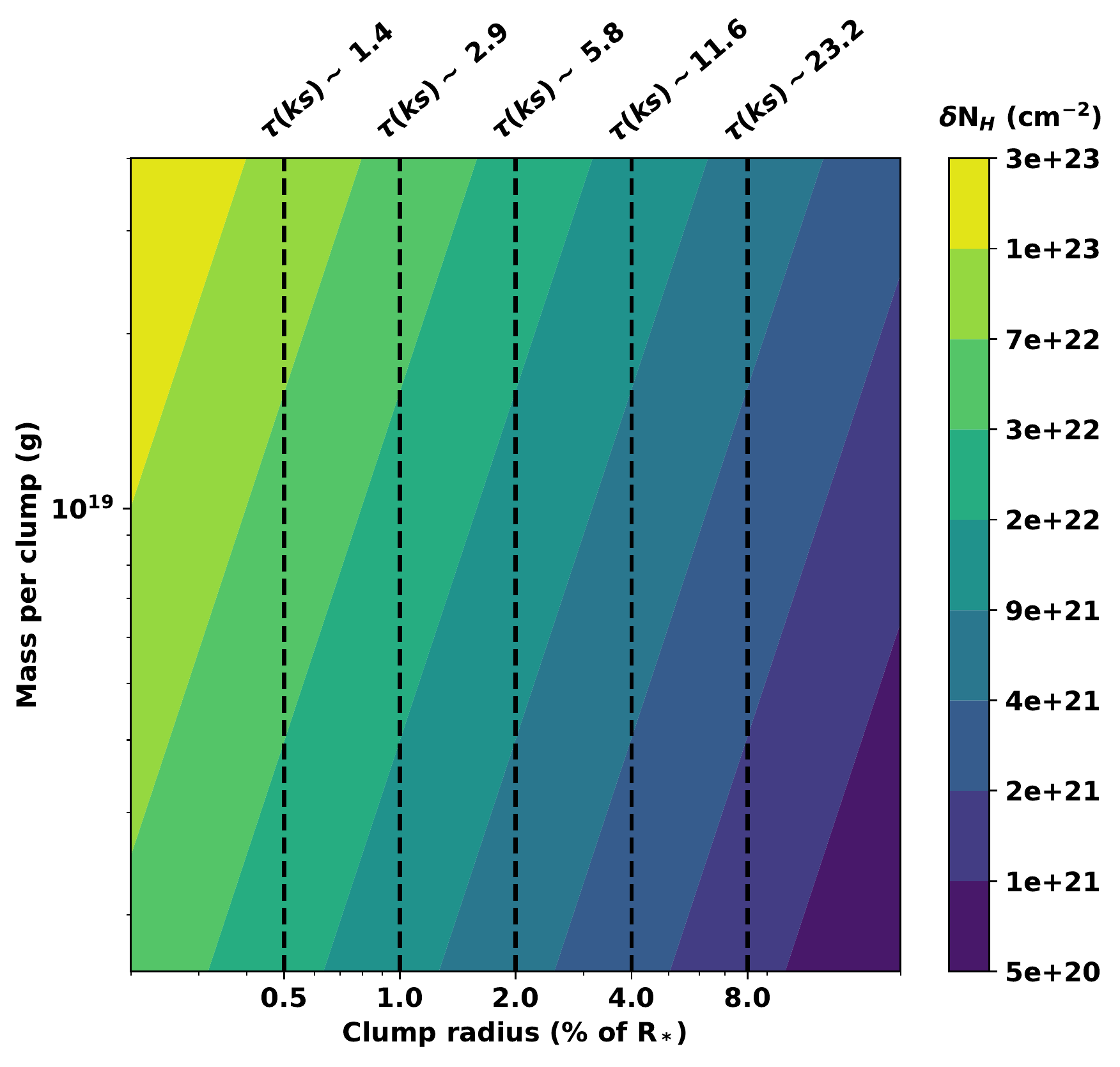}
\vspace{0.3cm}
\caption{Coherence time scale $\tau$ scale of column density time series is represented in ks at the top and the color map shows the standard deviation $\delta N_H$ for different clump sizes and masses. The values of orbital separation, the orbital inclination, eccentricity, longitude of periastron, and periastron angle are taken from \citet{J12} and the orbital phase of the pulsar is 0.25.}
\label{clumpsizes}
\end{figure*}
 
\subsection{Clump size estimates in \src: caveats and future outlook}
\label{sec:clumpsize}
{  


While X-ray variability is  attributed to clumpy stellar winds, there are a still a lot of open questions. For instance, although line driven instability can in principle generate important density variations in the stellar wind of massive stars (see e.g., \citealt{sundqvist2018}), the details and geometry are not yet fully understood. Furthermore, even if the wind is highly inhomogeneous, it is unclear whether the flow will be smoothed as it is accreted and clumps mix or whether additional instabilities will fragment the flow and make it more inhomogeneous. For instance, when the wind crosses the hydrodynamic bow shock surrounding the neutron star at a distance of the order of the accretion radius, the clumps mix in the shocked region \citep{elmellah2018}. It results in a smoother flow and a lower time variability than in the ballistic framework where clumps are instantaneously captured when their impact parameter is lower than the accretion radius \citep{ducci2009,oskinova2012}. On the other hand, magneto-centrifugal gating mechanisms at the outer rim of the neutron star magnetosphere could produce additional time variability \citep{bozzo2008}.

In the literature, there are discrepancies up to a factor of 10 on the clump sizes and of a factor of 1,000 on the clump masses \citep[see][]{pradhan2014, mellah2018}. At this point, hardness ratio-resolved X-ray spectroscopy is useful to constrain micro-structure of the wind like the density of individual clumps, the number of clumps per unit volume and the radius of clumps in HMXBs \citep{intzand2005_sfxts,pradhan2014,pradhan2019_igrj18027}.

Recently, \citet{elmellah2020} designed a model to characterize variable column density due to unaccreted foreground clumps intercepting the line-of-sight. The framework is built upon the first model drawn by \citet{grinberg2015}. It assumes that spherical clumps are launched from the photosphere of the donor star. The clumps expand as they flow away from the star on radial trajectories and follow a pre-defined beta velocity profile \citep{puls2008A}. The inter-clump environment is empty, the number of clumps is realistic and set by the stellar mass loss rate, the individual clump mass, the size of the system and the wind speed. They found that the statistical properties of the \nh~time series could be linked to the properties of the clumps. At each orbital phase, the characteristic time scale of absorption variability and the standard deviation with respect to the median column density profile depend on the size and on the mass of the individual clumps \citep[see][for more detail]{elmellah2020}. It is especially dependent on the properties of the clumps along the line of sight which are the closest from the donor star.

Taking advantage of the fact that OAO 1657$-$415 has been observed with numerous X-ray facilities like \suzaku~\citep{pradhan2014}, {\it Chandra}~\citep{oao_hetg2019}, {\it NICER} and \nustar~(this work) at various orbital phases, we attempt to constrain the clump sizes based on 
\citet{elmellah2020}. We computed the predicted coherence times  ($\tau$; time duration during which there is smooth variation in absorption column density) and standard deviation of the column density,  $\delta N_H$, for different clump masses and sizes of \src using the orbital parameters derived by \citet{falanga2015-orbit}. 

We first neglect the micro-structure of the wind and assume that it is smooth in order to derive the global parameters (i.e. the ones not linked to the clump properties). These parameters have been shown by \citet{elmellah2020} to be the orbital inclination $i$, the ratio of  the orbital separation $a$ to stellar radii $R_\star$ and the beta exponent $\beta$ (assuming a negligible eccentricity). In addition, the normalization of the profile, $N_{H,0}$, is  given by:
\begin{equation}
N_{H,0}=\frac{\dot{M}_\star}{R_\star v_{\infty}}
\end{equation}
where $\dot{M}_\star$ is the stellar mass loss rate and $v_{\infty}$ is the terminal wind speed. The best models we could obtain by qualitatively similar absorption profiles agree with the orbital parameters commonly derived in the literature: $i\sim$65--70 degrees and $a\sim 2R_\star$ \citep{falanga2015-orbit,mason2012}. Furthermore, the models favor a $\beta$ exponent around 2, suggestive of a progressive wind acceleration. The column density scales we find are of the order of 10$^{24}${\rm cm}$^{-2}$, coherent with a terminal wind speed of 250\,km s$^{-1}$, a stellar radius of 25R$_{\odot}$ and a mass loss rate of 1$\times10^{-6}$ M$_{\odot}$ yr$^{-1}$, though the latter is somewhat higher than the upper limit reported by \citet{falanga2015-orbit}. We could then constrain the clump size and mass based on Figure 10 in \citet{elmellah2020} but for parameters suited for \src. In Figure \ref{clumpsizes}, the coherence time scale $\tau$ is represented in kiloseconds at the top while the color map shows the standard deviation $\delta N_H$ for different clump sizes and masses for orbital phase 0.25. This exercise illustrates a novel methodology to measure more accurately different clump sizes and masses in HMXBs.

The above method is particularly important since the fundamental prerequisite to characterize clumps in stellar wind is to have the ability to makes measurements in absorption at timescales of the order of or smaller than the coherence timescale. However, the limitation set by the detector properties (area, cadence of observations) hinders the study of time-resolved spectroscopy down to the required short time scales and therefore the clump size estimates determined this way should be regarded as upper limits. Therefore, for example, from \suzaku~measurements, we could measure the absorption changes in timescales of only $\sim$ 10\,ks (Fig.~9 of \citealt{pradhan2014}), the relatively small collecting area of the {\it Suzaku}/XIS do not permit time-resolved spectroscopy at shorter time scales. The fast spectral changes associated with the source can be most effectively studied with \xmm~due to its ability to observe the source uninterruptedly, thanks to the orbit of the satellite\footnote{\xmm~has not observed this source so far.}. The other two operating X-ray spacecrafts with largest areas at the present are \nustar~and {\it NICER}, the results of which are presented here. While both are indispensable in investigating the X-ray properties of the pulsar as we see above, both have limitations caused by cadence of observation. The low altitude and low inclination orbit of \nustar~are inter dispersed with gaps in data thereby limiting the required cadence necessary for this kind of work (see section 6.2 of \citealt{elmellah2020} for details). On the other hand, while NICER has been very efficient in orbital monitoring of the source spread across days, the cadence of one observation can be less than 90\,min long over which the spacecraft orbits Earth.

Short-term variability studies of X-ray binaries on smaller time scales have been made in the literature using indirect methods such color-color diagrams \citep{nowak2011,grinberg2020}. In the near future, high throughput X-ray missions such as {\it XRISM} \citep{xrism2020}, {\it eXTP} \citep{intzand2019}, and {\it Athena} \citep{nandra2013} can enable short-term absorption variability studies, even for moderately faint sources. With {\it Athena}, for example, high S/N spectra of HMXBs are possible in timescales of few hundred seconds \citep{lomaeva2020,Amato2021}. Therefore, efforts like the current paper - dedicated to the study of short term variability of absorption and their implications - are now necessary to act as segues for these upcoming high-resolution missions and their role in studying the inhomogeneity of stellar winds. 
}

\section{Summary}
\label{summary}

In this paper we exploited the \nustar~and \nicer~observations of \src in order to carry out for the first time a detailed HR-resolved spectral analysis of the X-ray emission from \src. Taking advantage of the good energy resolution and large effective area of \nustar, the HR-resolved spectral analysis could be carried out in all cases on few ks timescales and thus be used to characterize the clumpy winds and locate where emission lines are formed in \src. Additionally, data from both instruments helped us characterize the pulse profiles in \src. We summarize the main findings of our work below.

\begin{itemize}

\item The HR and spin phase resolved spectroscopic studies of \src~with \nustar~and the spectral variation obtained by monitoring the source with \nicer~are consistent with the presence of clumpy stellar winds in the system. 

\item We report for the first time a dip around 0.15 in the pulse profiles from \nustar, possibly caused by changes in viewing geometry. Probable explanations include X-ray emission in multiple accretion columns or spectral changes due to the formation of a transient disk.

\item The neutral iron lines are formed far away from the neutron star - possibly in the clumps in the stellar wind while the ionized lines are formed closer to the neutron star and probably within the magnetosphere. 

\item The \nustar~spectrum clearly revealed that the putative CRSF at $\sim$ 36\,keV is not a real feature and the dip reported in the literature is possibly caused by turnover of the X-ray spectrum at $\sim$ 20 \,keV \citep{O99}. 

\item We did not detect pulsations in 2 of the 6 \nicer~observations. This is possibly due to increased absorption of X-ray photons in the surrounding matter. 

\item We illustrate a novel methodology by \citet{elmellah2020} to measure clump sizes and masses in HMXBs more accurately based on absorption measurements and orbital parameters of the source. Finally, we segue our current work into the future of X-ray binaries with upcoming X-ray missions. 
\end{itemize}

\clearpage

\section*{Acknowledgements}
This research has made use of data and software provided by the High Energy Astrophysics Science Archive Research Center (HEASARC), which is a service of the Astrophysics Science Division at NASA/GSFC and the High Energy Astrophysics Division of the Smithsonian Astrophysical Observatory. PP would like to acknowledge NASA \nustar~grant award number 029926-00001 and \nicer~opportunity NNH19ZDA001N for partially supporting this work. 

{  
\appendix
\label{appendix}
\renewcommand{\thefigure}{A\arabic{figure}}
\counterwithin{figure}{section}
\section{Corner plots}

The appendix contains corner plots for the average spectrum and three hardness-ratio resolved and three pulse phase resolved segments where the absorption column density are low, medium and high.

\begin{figure}[htb]
  \centering\leavevmode
  \includegraphics[width=1.0\columnwidth,angle=0]{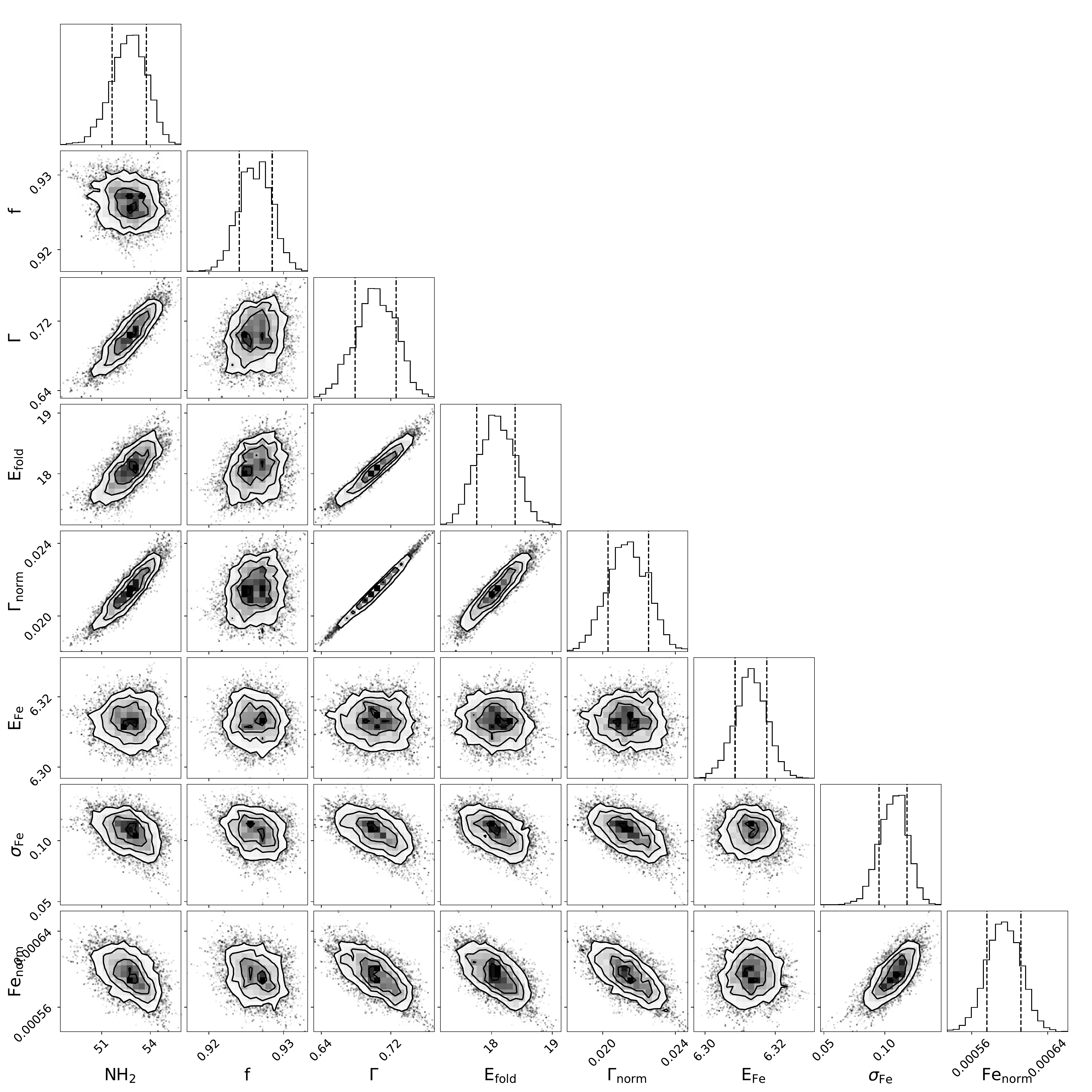}
  \caption{Corner plots for the average spectrum. }
  \label{corner:avg}
\end{figure}


\begin{figure}[htb]
  \centering\leavevmode
   \includegraphics[width=1.0\columnwidth,angle=0]{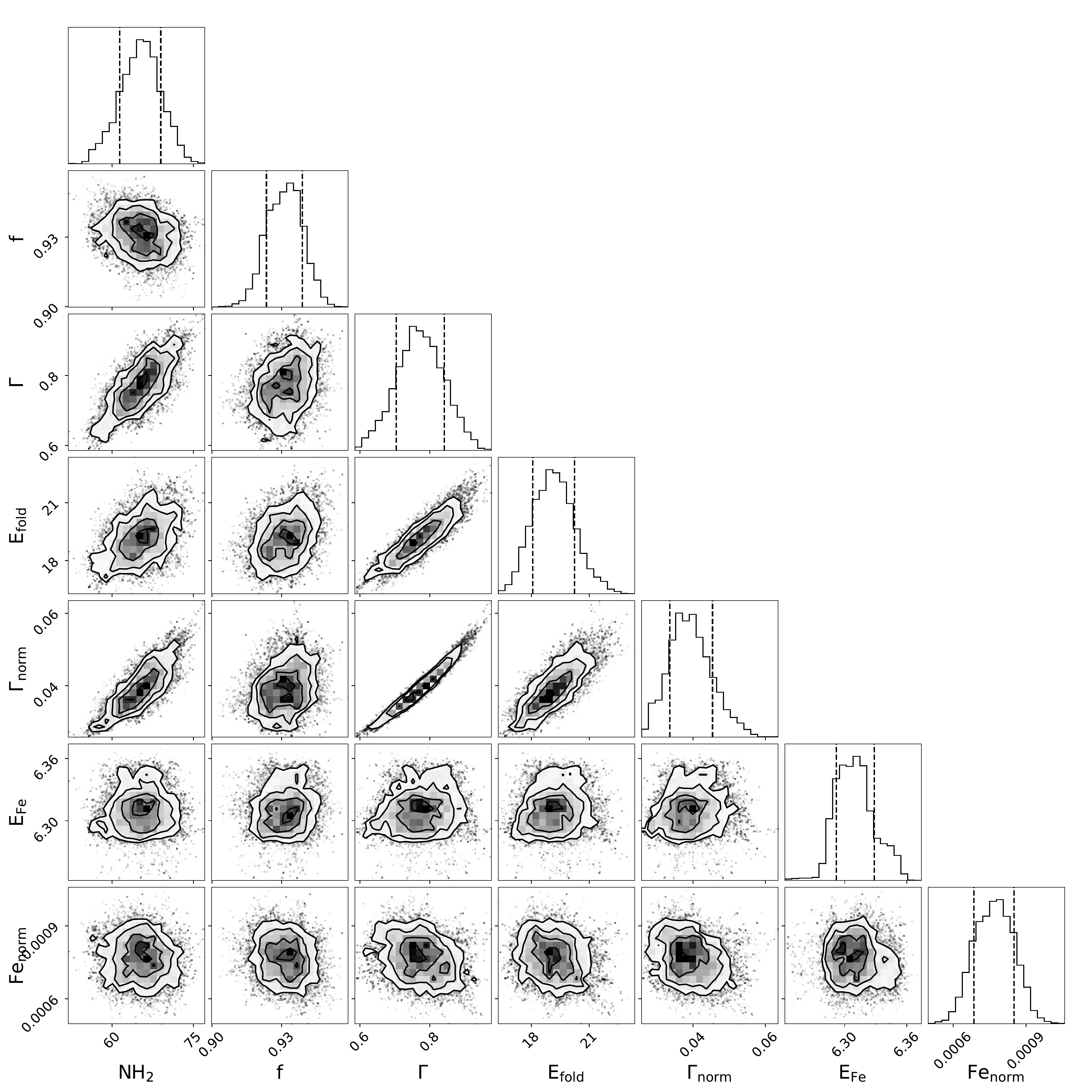}
  \caption{Corner plots for hardness-ratio resolved segment 1. }
  \label{corner:hr1}
\end{figure} 

\begin{figure}[htb]
  \centering\leavevmode
   \includegraphics[width=1.0\columnwidth,angle=0]{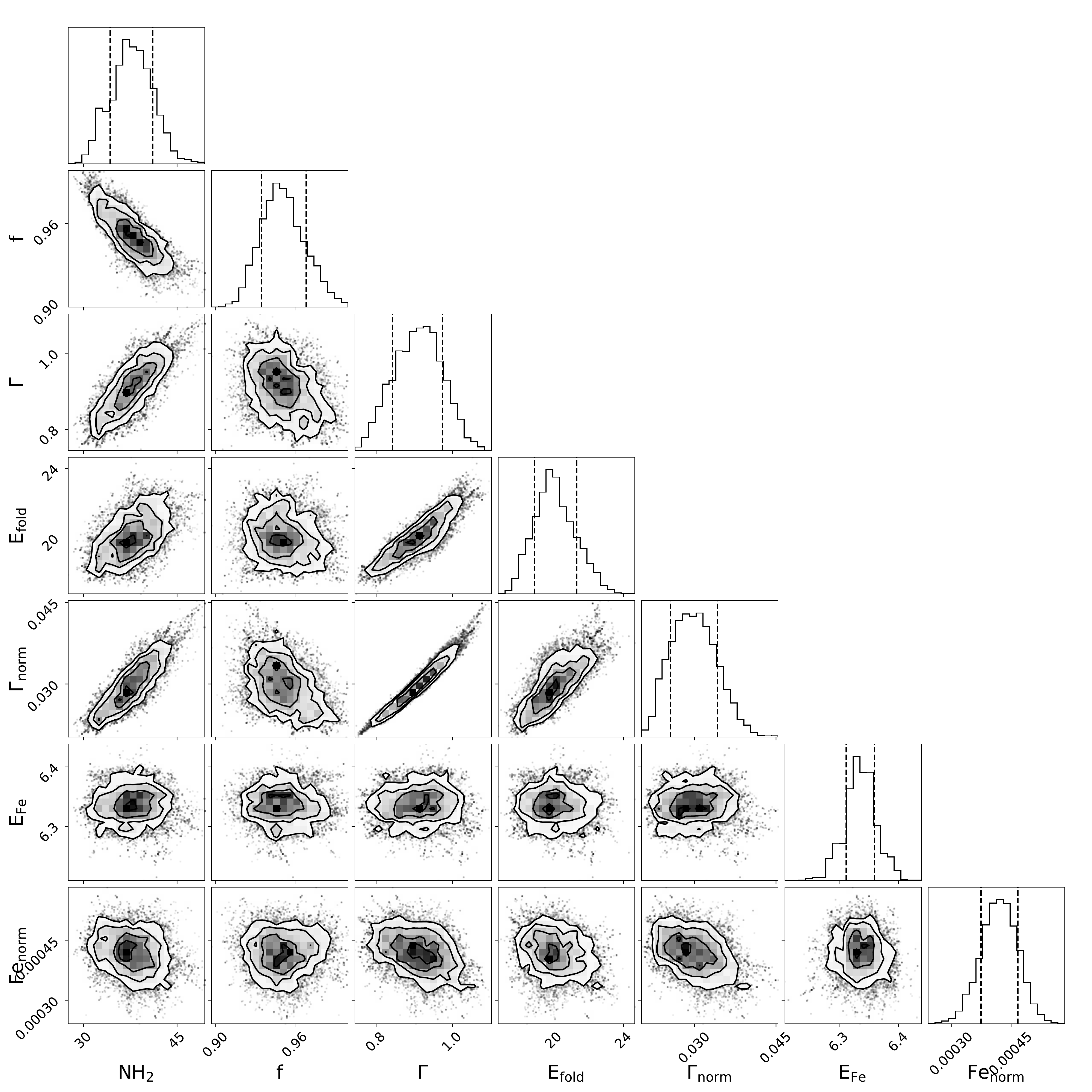}
  \caption{Corner plots for hardness-ratio resolved segment 7.}
  \label{corner:hr2}
\end{figure} 

\begin{figure}[htb]
  \centering\leavevmode
  \includegraphics[width=1.0\columnwidth,angle=0]{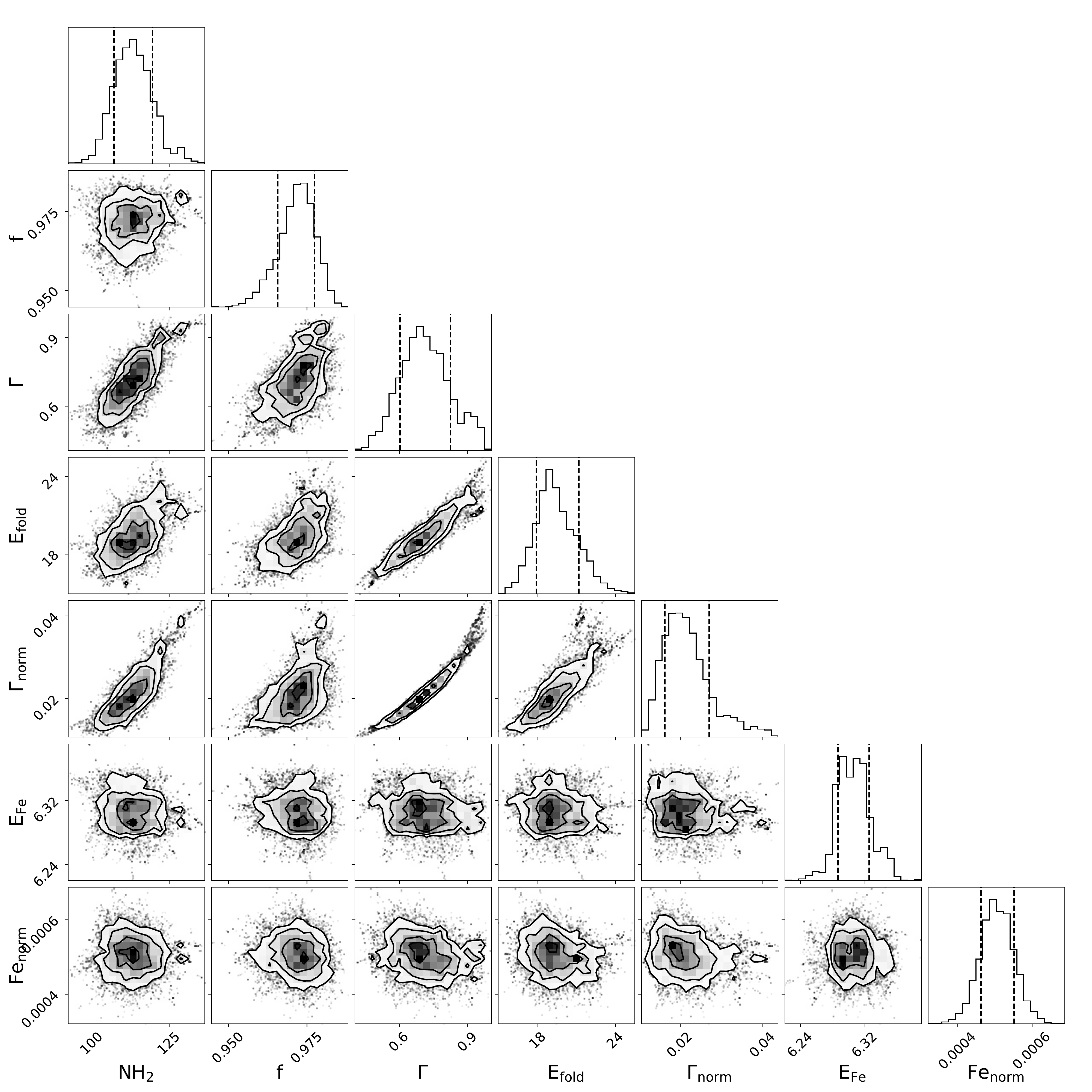}
  \caption{Corner plots for hardness-ratio resolved segment 19.}
  \label{corner:hr3}
\end{figure}


\begin{figure}[htb]
  \centering\leavevmode
  \includegraphics[width=1.0\columnwidth,angle=0]{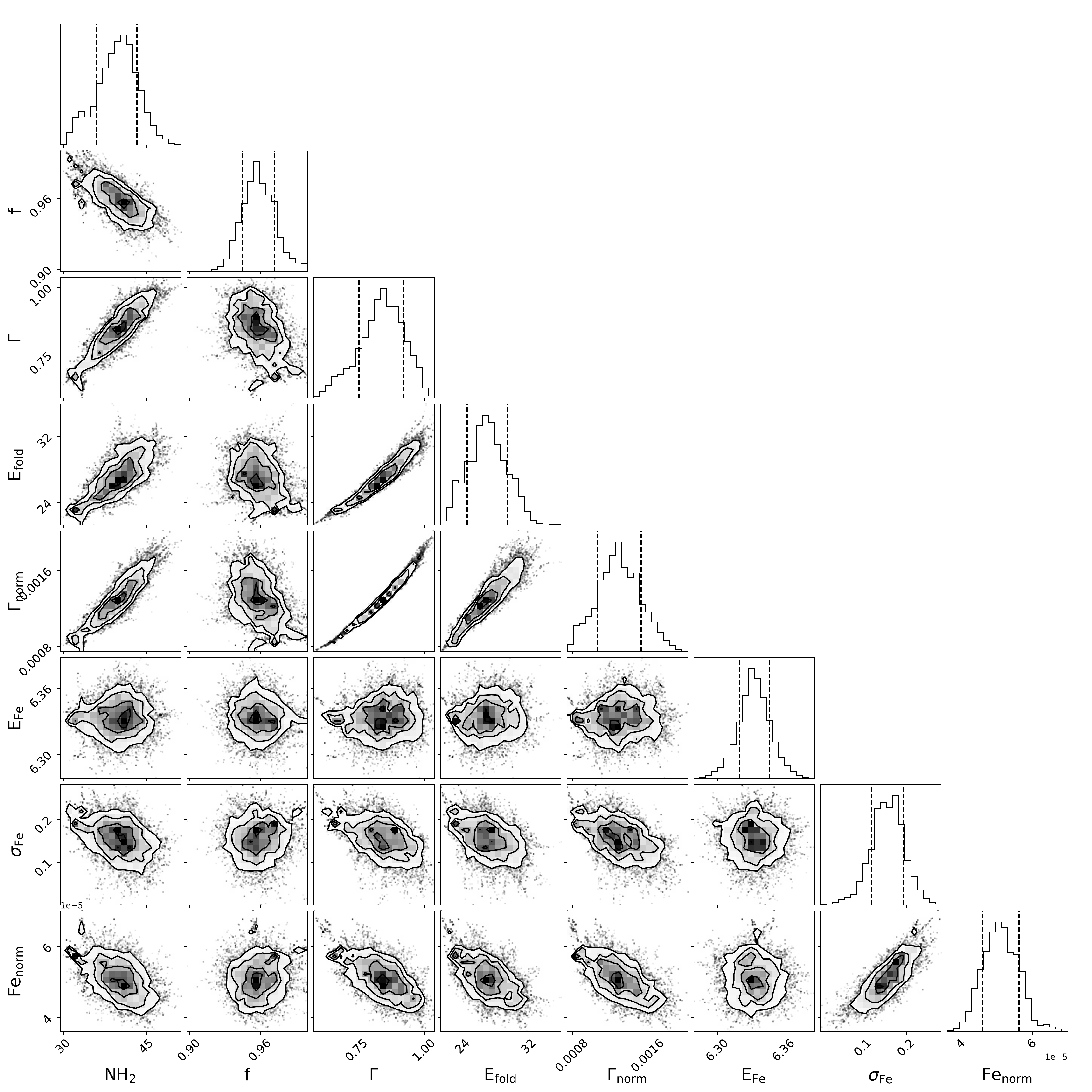}
  \caption{Corner plots for third phase bin (phase $\sim$ 0.2) in Fig.~\ref{spec-phase}. }
  \label{corner:pr1}
\end{figure}

\begin{figure}[htb]
  \centering\leavevmode
  \includegraphics[width=1.0\columnwidth,angle=0]{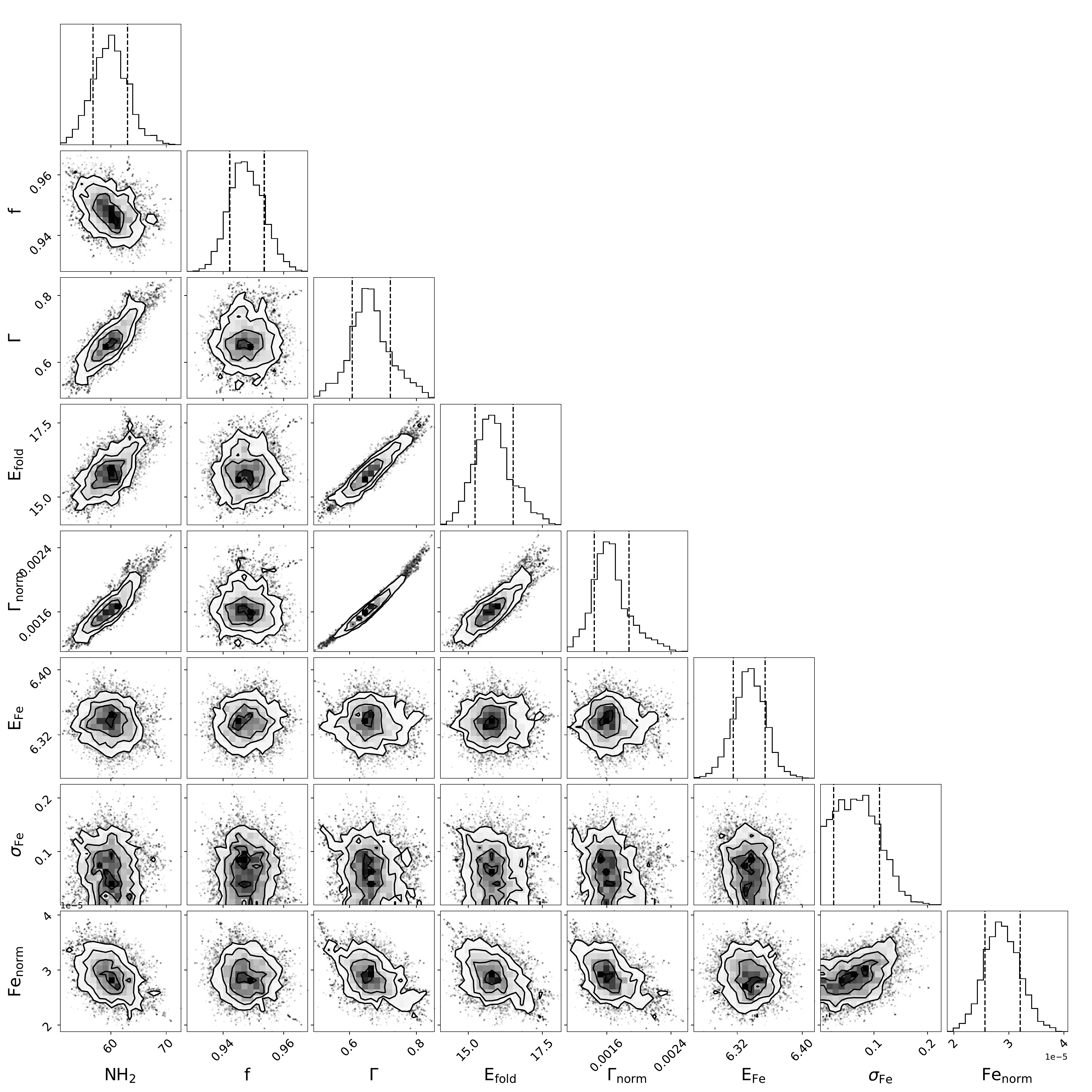}
  \caption{Corner plots for seventh phase bin (phase $\sim$ 0.4) in Fig.~\ref{spec-phase}. }
  \label{corner:pr2}
\end{figure}

\begin{figure}[htb]
  \centering\leavevmode
  \includegraphics[width=1.0\columnwidth,angle=0]{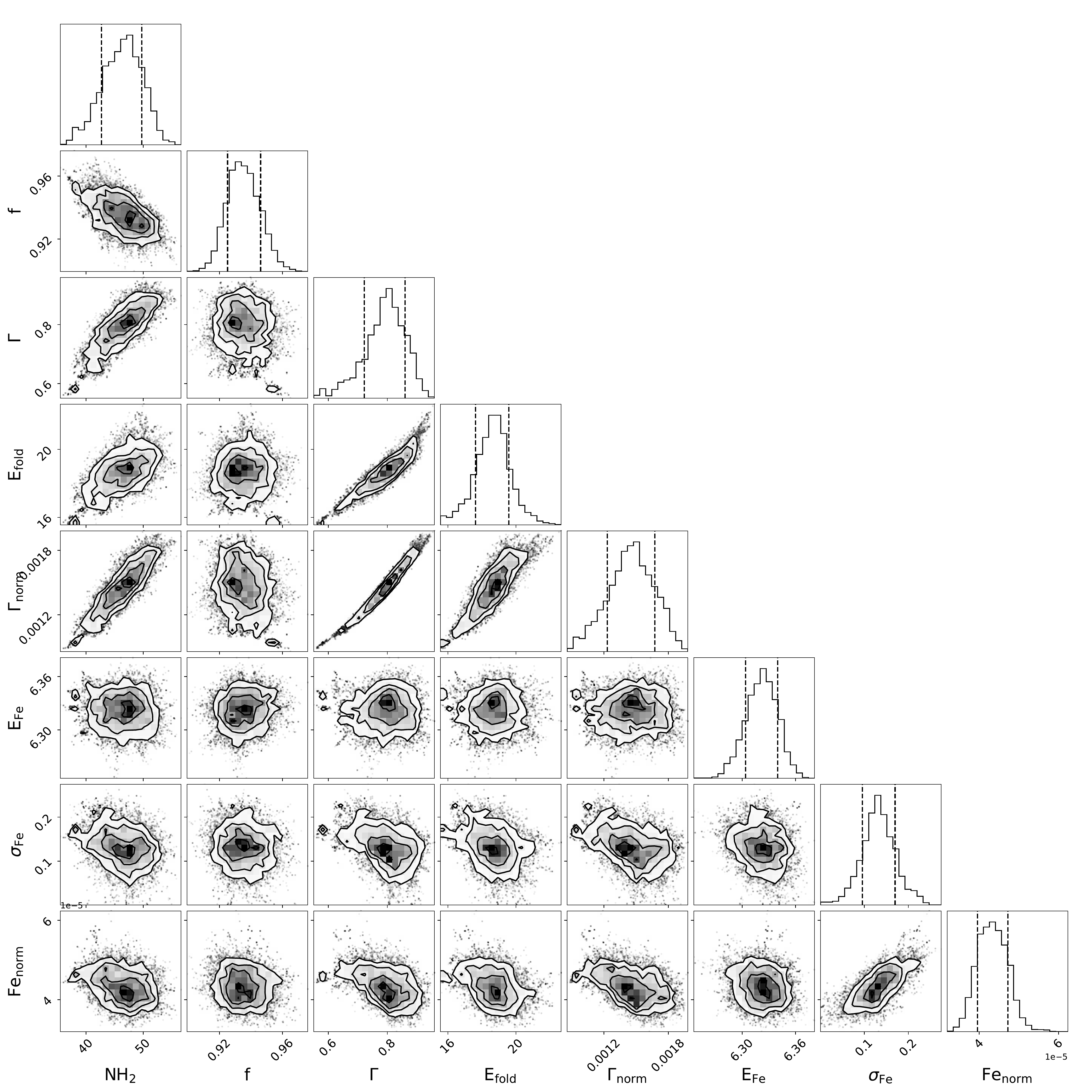}
  \caption{Corner plots for thirteenth phase bin (phase $\sim$ 0.8) in Fig.~\ref{spec-phase}.}
  \label{corner:pr3}
\end{figure}

}

\bibliography{ref_short}


\end{document}